\documentclass[twocolumn]{aastex63}

\usepackage{natbib}
\usepackage{multirow}
\usepackage{color}
\usepackage{bm}
\usepackage[normalem]{ulem}

\newcommand{\blue}{\textcolor{black}}
\newcommand{\green}{\textcolor{black}}
\newcommand{\greenn}{\textcolor{black}}

\received{\today}
\revised{\today}
\accepted{\today}
\submitjournal{ApJ}

\shorttitle{Rapidly Evolving Transients from the Hyper Suprime-Cam SSP Transient Survey}
\shortauthors{Y. Tampo et al.}
\begin{document}

\title{Rapidly Evolving Transients from the Hyper Suprime-Cam SSP Transient Survey}

\correspondingauthor{Yusuke Tampo}
\email{tampo@kusastro.kyoto-u.ac.jp}

\author{Yusuke Tampo}
\affiliation{Department of Astronomy, Kyoto University, Kitashirakawa-Oiwake-cho, Sakyo-ku, Kyoto 606-8502, Japan}
\affiliation{Astronomical Institute, Tohoku University, Aoba, Sendai 980-8578, Japan}

\author{Masaomi Tanaka}
\affiliation{Astronomical Institute, Tohoku University, Aoba, Sendai 980-8578, Japan}
\affiliation{Kavli Institute for the Physics and Mathematics of the Universe (WPI), The University of Tokyo Institutes for Advanced Study, The University of Tokyo, 5-1-5 Kashiwanoha, Kashiwa, Chiba 277-8583, Japan}

\author{Keiichi Maeda}
\affiliation{Department of Astronomy, Kyoto University, Kitashirakawa-Oiwake-cho, Sakyo-ku, Kyoto 606-8502, Japan}
\affiliation{Kavli Institute for the Physics and Mathematics of the Universe (WPI), The University of Tokyo Institutes for Advanced Study, The University of Tokyo, 5-1-5 Kashiwanoha, Kashiwa, Chiba 277-8583, Japan}

\author{Naoki YASUDA}
\affiliation{Kavli Institute for the Physics and Mathematics of the Universe (WPI), The University of Tokyo Institutes for Advanced Study, The University of Tokyo, 5-1-5 Kashiwanoha, Kashiwa, Chiba 277-8583, Japan}

\author{Nozomu TOMINAGA}
\affiliation{Department of Physics, Faculty of Science and Engineering, Konan University, 8-9-1 Okamoto, Kobe, Hyogo 658-8501, Japan}
\affiliation{Kavli Institute for the Physics and Mathematics of the Universe (WPI), The University of Tokyo Institutes for Advanced Study, The University of Tokyo, 5-1-5 Kashiwanoha, Kashiwa, Chiba 277-8583, Japan}

\author{Ji-an JIANG}
\affiliation{Kavli Institute for the Physics and Mathematics of the Universe (WPI), The University of Tokyo Institutes for Advanced Study, The University of Tokyo, 5-1-5 Kashiwanoha, Kashiwa, Chiba 277-8583, Japan}

\author{Takashi J. MORIYA}
\affiliation{National Astronomical Observatory of Japan, National Institutes of Natural Sciences, 2-21-1 Osawa, Mitaka, Tokyo 181-8588, Japan}
\affiliation{School of Physics and Astronomy, Faculty of Science, Monash University, Clayton, VIC 3800, Australia}

\author{Tomoki MOROKUMA}
\affiliation{Institute of Astronomy, Graduate School of Science, The University of Tokyo, 2-21-1 Osawa, Mitaka, Tokyo 181-0015, Japan}
\affiliation{Kavli Institute for the Physics and Mathematics of the Universe (WPI), The University of Tokyo Institutes for Advanced Study, The University of Tokyo, 5-1-5 Kashiwanoha, Kashiwa, Chiba 277-8583, Japan}

\author{Nao SUZUKI}
\affiliation{Kavli Institute for the Physics and Mathematics of the Universe (WPI), The University of Tokyo Institutes for Advanced Study, The University of Tokyo, 5-1-5 Kashiwanoha, Kashiwa, Chiba 277-8583, Japan}

\author{Ichiro TAKAHASHI}
\affiliation{Kavli Institute for the Physics and Mathematics of the Universe (WPI), The University of Tokyo Institutes for Advanced Study, The University of Tokyo, 5-1-5 Kashiwanoha, Kashiwa, Chiba 277-8583, Japan}
\affiliation{CREST, JST, 4-1-8 Honcho, Kawaguchi, Saitama 332-0012, Japan}

\author{Mitsuru KOKUBO}
\affiliation{Astronomical Institute, Tohoku University, Aoba, Sendai 980-8578, Japan}

\author{Kojiro KAWANA}
\affiliation{Department of Physics, School of Science, The University of Tokyo, 7-3-1, Bunkyo, Tokyo 113-0033, Japan}



\begin{abstract}

Rapidly evolving transients form a new class of transients which show shorter timescales of the light curves than those of typical core-collapse and thermonuclear supernovae. We performed a systematic search for rapidly evolving transients using the deep data taken with the Hyper Suprime-Cam Subaru Strategic Program Transient Survey. By measuring the timescales of the light curves of 1824 transients, we identified 5 rapidly evolving transients. Our samples are found in a wide range of redshifts (0.3 $\le$ z $\le$ 1.5) and peak absolute magnitudes ($-$17 $\ge$ $M_i$ $\ge$ $-$20). The properties of the light curves are similar to those of the previously discovered rapidly evolving transients.
They show a relatively blue spectral energy distribution, with the best-fit blackbody of 8,000 - 18,000 K. We show that some of the transients require power sources other than the radioactive decays of $^{56}$Ni because of their high peak luminosities and short timescales. The host galaxies of all the samples are star-forming galaxies, suggesting a massive star origin for the rapidly evolving transients. The event rate is roughly estimated to be $\sim$4,000 events yr$^{-1}$ Gpc$^{-3}$, which is about 1 $\%$ of core-collapse supernovae.

\end{abstract}


%\keywords{editorials, notices --- miscellaneous --- catalogs --- surveys}
\keywords{supernovae: general}

\section{Introduction}
\label{sec:intro}
Over the past decades, various types of supernovae (SNe) have been studied, and general understanding of progenitors and explosion scenarios for normal types of SNe has been established: Type Ia SNe (SNe Ia) are thermonuclear explosions of white dwarfs while SNe II and SNe Ibc are explosions caused by core-collapse of massive stars ($>$8 M$_{\odot}$). However, recent dedicated time-domain surveys with various depths (e.g., Sloan Digital Sky Survey, \citealt{SDSS}; Palomar Transient Factory, \citealt{PTF_1, PTF_2}; Pan-STARRS1 (PS1), \citealt{kaiser10}; ATLAS, \citealt{atlas}, and HST/CANDLES, \citealt{candles}) are discovering an increasing number of SNe, and revealing the populations of rare classes of SNe showing unique properties, such as super-Chankdrasekhar SNe, \citep{SC-SN} or superluminous SNe (SLSNe; \citealt{SLSN_found}).

A class of rapidly evolving transients is one of such peculiar types of explosive phenomena.
Their light curves evolve much faster than the normal SNe. Until today, a considerable number of rapidly evolving transients have been identified, including SN 2002bj \citep{2002bj}, SN 2005ek \citep{2005ek}, SN 2010X \citep{2010X}, iPTF 16asu from Palomor Transient Factory \citep{iPTF_16asu}, KSN2015K from Kepler mission \citep{KSN2015K}, SN 2017czd \citep{2017czd}, AT2018cow \citep{COW, COW_BH}, SN 2018bgv \citep{SN18bgv}, SN 2018gep \citep{2018gep}, SN 2019 bkc \citep{2019bkc_a,2019bkc_b}, SN 2018kzr \citep{sn2018kzr} and SHOOT14di \citep{tom_rapid_decline}. 
Also, a few systematic searches have been performed and 
14 and 72 rapidly evolving transients have been identified with PS1 (\citealt{PS1}) and Dark Energy Survey (DES, \citealt{DES}), respectively.

The light curves of these rapidly evolving transients evolve faster than those of normal SNe both in the rising and declining phases.
Their luminosities show a large diversity, ranging from absolute magnitudes of $-$15 to $-$22 mag. 
Their progenitors and power sources have not yet been  clarified. 
For example, radioactive decays of $^{56}$Ni, the primary power source of many types of SNe, cannot reproduce the light curves of luminous rapidly evolving transients \citep{PS1, DES}. Also, although some spectroscopic data are taken for rapidly evolving transients, some of them only show blue continuum and a lack of strong absorption/emission lines.
Interestingly, the event rate is not negligible among all types of SNe: it has been estimated as 4,800 - 8,000 events yr$^{-1}$ Gpc$^{-3}$ \citep{PS1}, which corresponds to about 1$\%$ of that of core-collapse SNe.
 
There are some suggestions for the explosion scenarios of rapidly evolving transients, including transients associated with shock breakout from massive stars \citep{PS1}, or from circumstellar material (CSM, \citealt{PS1, KSN2015K, arcavi_rapid}), ultra-stripped envelope SNe \citep{2013ultra_strip, ultra_stripped}, electron-capture SNe \citep{electron_capture}, magnetar-powered SNe \citep{iPTF_16asu, arcavi_rapid}, failed core-collapse SNe \citep{failed_ccsn}, explosions of extended Wolf-Rayet stars \citep{extended_WR}, tidal disruption events \citep{PS1, kawana2019rapid} and peculiar thermonuclear explosions \citep{thermonuclear_explo}. 
Since rapidly evolving transients have a large variety in their timescales and peak luminosities, they may have multiple origins \citep{PS1, DES}.

In this paper, we present our search for rapidly evolving transients using the data from
the Hyper Suprime-Cam Subaru Strategic Program Transient Survey (HSC-SSP Transient Survey). Compared with the past surveys, the HSC-SSP Transient Survey is deeper, which enables us to detect objects at high redshifts, and covers the ultraviolet (UV) wavelengths in the rest frame. 

This paper is structured as follows. In Section $\ref{sec:selection}$, we give a brief overview of the HSC-SSP Transient Survey and our criteria for sample selection. We present our samples in Section $\ref{sec:samples}$, and then show photometric properties in Section $\ref{sec:photometric}$. In Section $\ref{sec:discussion}$,
we show properties of the host galaxies of our samples and discuss the event rate of rapidly evolving transients as well as possible power sources. In Section $\ref{sec:conclusions}$, we give conclusions of our paper.
Throughout of the paper, the magnitude is given in the AB magnitude system.
All calculations in this paper assume Planck13 model; a flat $\Lambda$CDM cosmology with H0 = 67.8 km s$^{-1}$ Mpc$^{-1}$, $\Omega_m$ = 0.307 \citep{Planck13}.

\section{Observations and sample selection}
\label{sec:selection}

\subsection{HSC-SSP Transient Survey}
We use the observational data from the HSC-SSP Transient Survey \citep{hsc_basic, HSC_survey}.
The survey around the COSMOS field was performed over 6 months from November 2016 to April 2017, covering the 1.77 deg$^2$ Ultra-Deep layer and the 5.78 deg$^2$ Deep layer, with the cadence of 7-10 days using the $g$-, $r$-, $i$-, $z$- and $y$-bands \citep{HSC_survey}. Note that the $y$-band data suffer from scattered lights \citep{hsc_basic}, and therefore, we do not use the $y$-band data for our analysis as in \citet{HSC_survey}.
Limiting magnitude in each epoch was estimated by detecting and measuring artificially injected point sources into the CCD images 
(see \citealt{HSC_survey} for more details).
Derived median depths per epoch are 26.4, 26.3, 26.0 and 25.6 mag in the $g$-, $r$-, $i$- and $z$-bands for the Ultra-Deep layer, and $\sim$0.6 mag shallower for the Deep layer. 
Compared to the past surveys such as PS1 and Dark Energy Survey, 
the HSC-SSP Transient Survey adopts longer cadence and a fewer number of filters on the same epoch,
but it is much deeper. This better sensitivity enables us to find objects at high redshifts.

In this survey, 1824 SN candidates were identified by \citet{HSC_survey}. 
\blue{To identify the host galaxies of the candidates, the closest object in the reference images was selected as the host object. At this stage, clear misidentifications were corrected by visual inspection \citep{HSC_survey}. Then, the host objects were matched in a search radius of 0.5 arcsec with SDSS DR12 \citep{SDSS_DR12}, PRIMUS DR1 \citep{PRIMUS1, PRIMUS2}, VVDS \citep{VVDS_DR}, zCOSMOS DR3 \citep{zCOSMOS_DR}, 3D-HST \citep{3D0HST_five, 3D-HST_2}, COSMOS2015 \citep{laigle16} and HSC photo-z catalog \citep{HSC_photoz}.}

Through the template fitting of SALT2 SN Ia light curves \citep{guy07, Ia_rel}, 433 objects out of 1824 SN candidates were labeled as genuine SNe Ia. 
It is noted that, since the purpose of the template fitting in \citet{HSC_survey} is to identify genuine SNe Ia with standard properties, the remaining 1391 objects may still include SNe Ia. 
We therefore perform multi-type template fitting for all the 1824 objects by using sncosmo package \citep{barbary16}. 
\blue{
For SNe Ia, we use the SALT2 template \citep{guy07}.  We varied the stretch parameter ($x1$), the color parameter ($c$), the maximum brightness, and the peak time as free parameters. For SNe II and SNe Ibc, we use 27 and 15 templates constructed from the SNANA package \citep{kessler09}, respectively, to take into account the variety of the light curves. We varied the maximum brightness and the peak time as free parameters. The extinction is not included for SNe II and Ibc fitting. Finally, the best fit $\chi^2$ value for each type are used to select candidates of rapidly evolving transients (see the next section for our criteria).}

\subsection{Selection Criteria for Rapidly Evolving Transients}

\begin{figure}[htbp]
 \begin{center}
  \includegraphics[width=81mm]{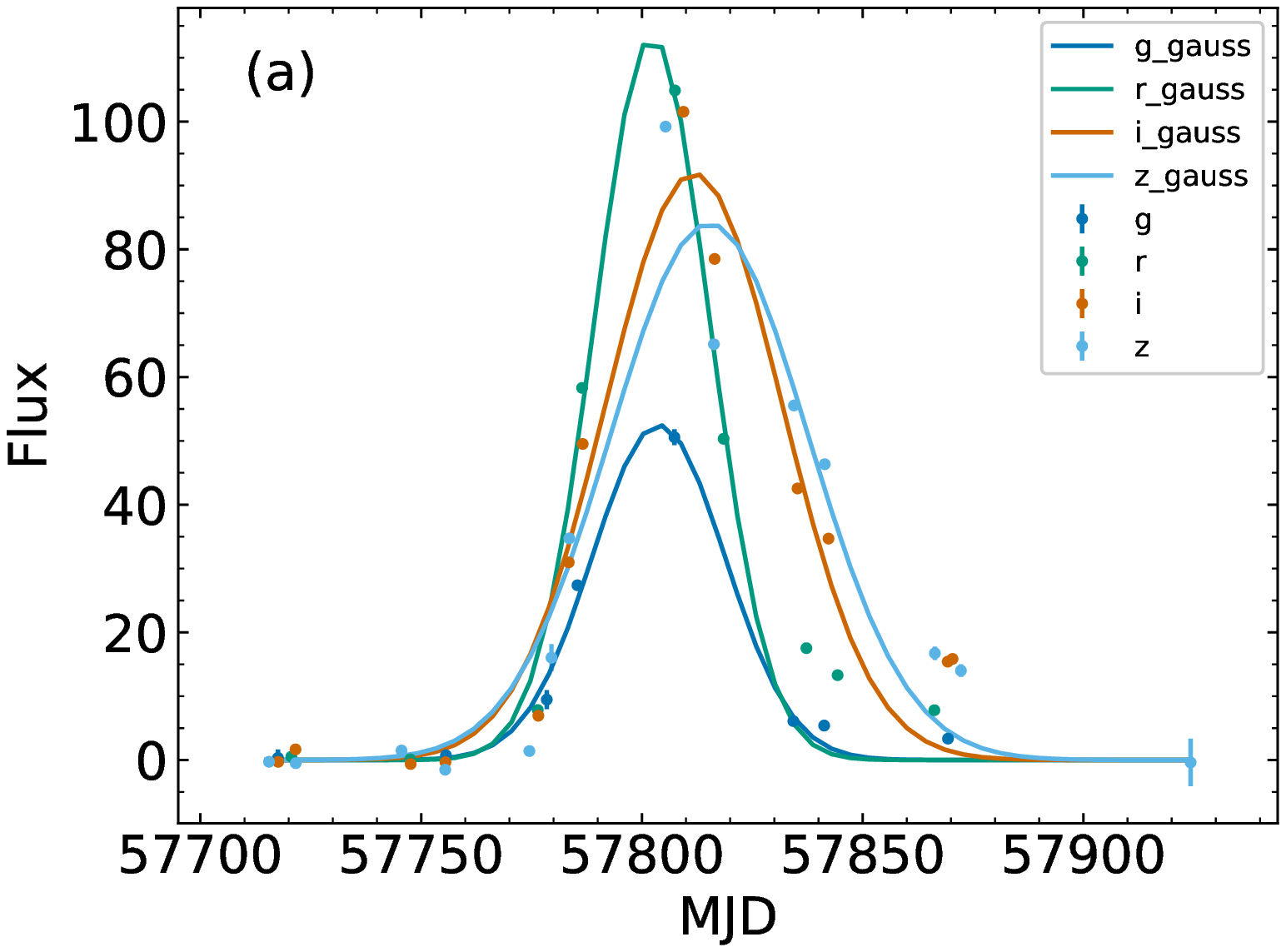}
  \includegraphics[width=81mm]{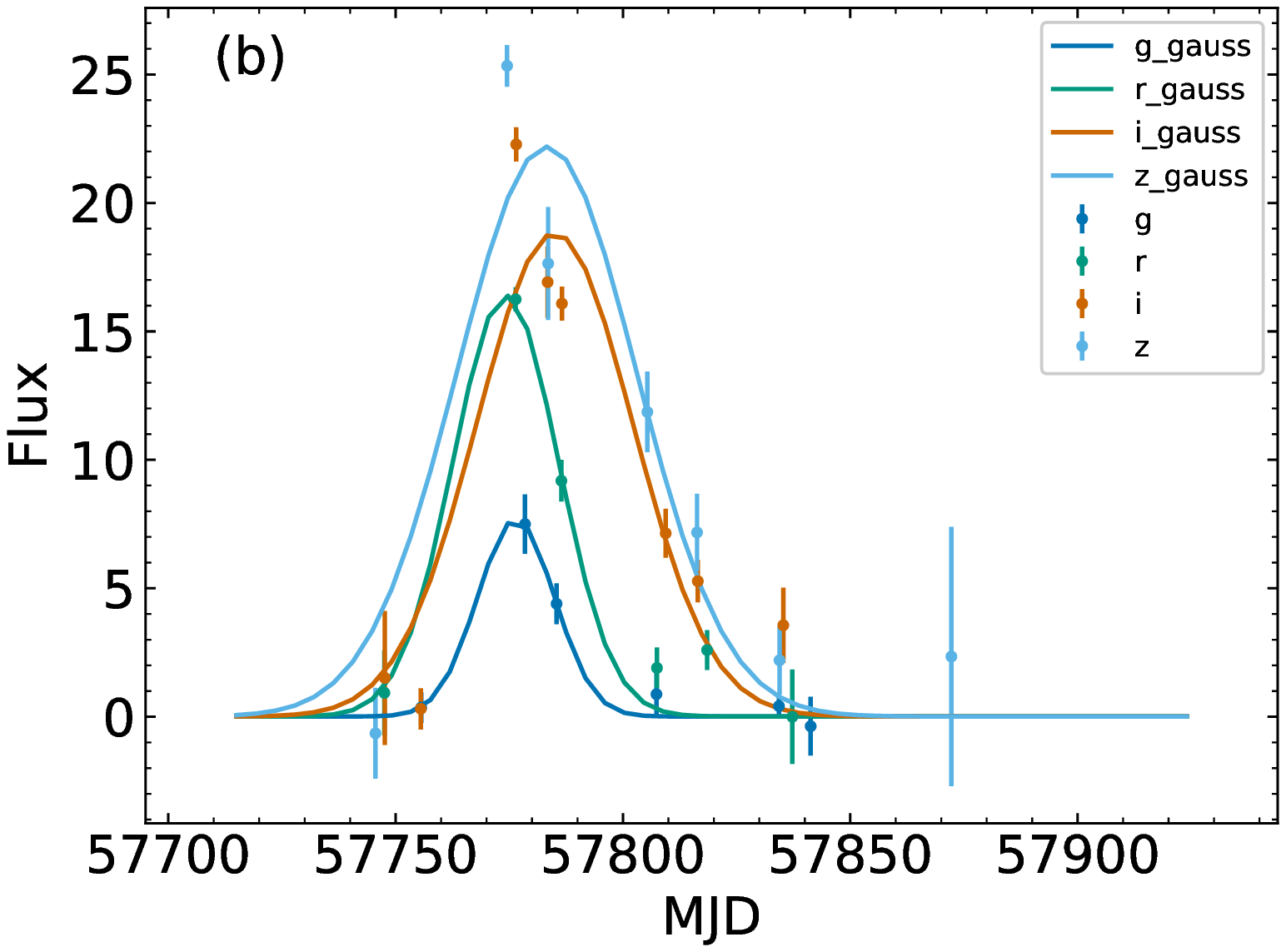}
 \end{center}
 \caption{Examples of the observed light curves for a transient that is classified as a genuine SN Ia (HSC17bmhk; (a)) and a transient that is not classified as a genuine SN Ia (HSC17bgji; (b)). The solid lines show the best fit Gaussian functions.}
  \label{fig:gaussianedsample}
\end{figure}

\begin{figure}[htbp]
 \begin{center}
  \includegraphics[width=80mm]{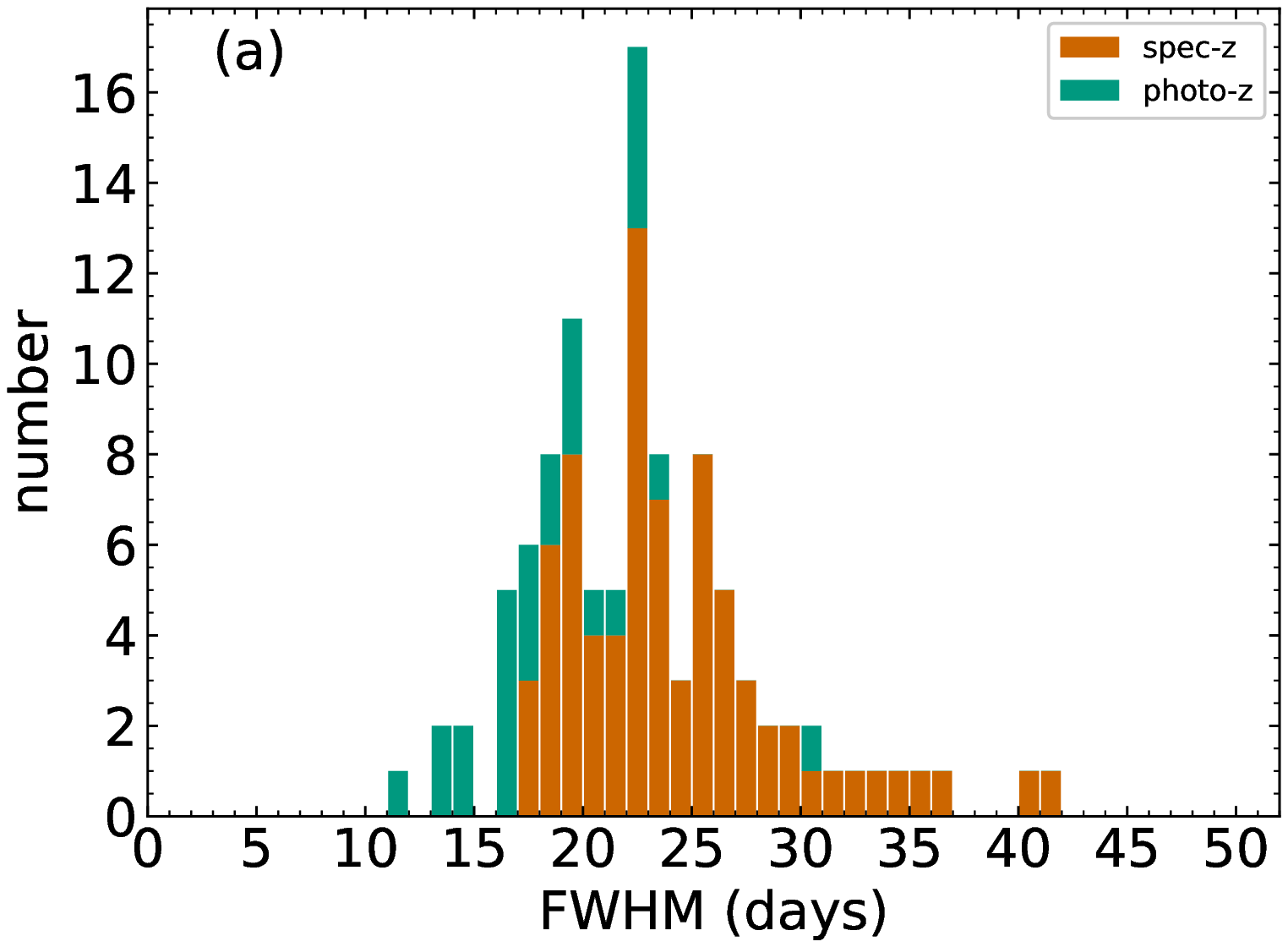}
  \includegraphics[width=80mm]{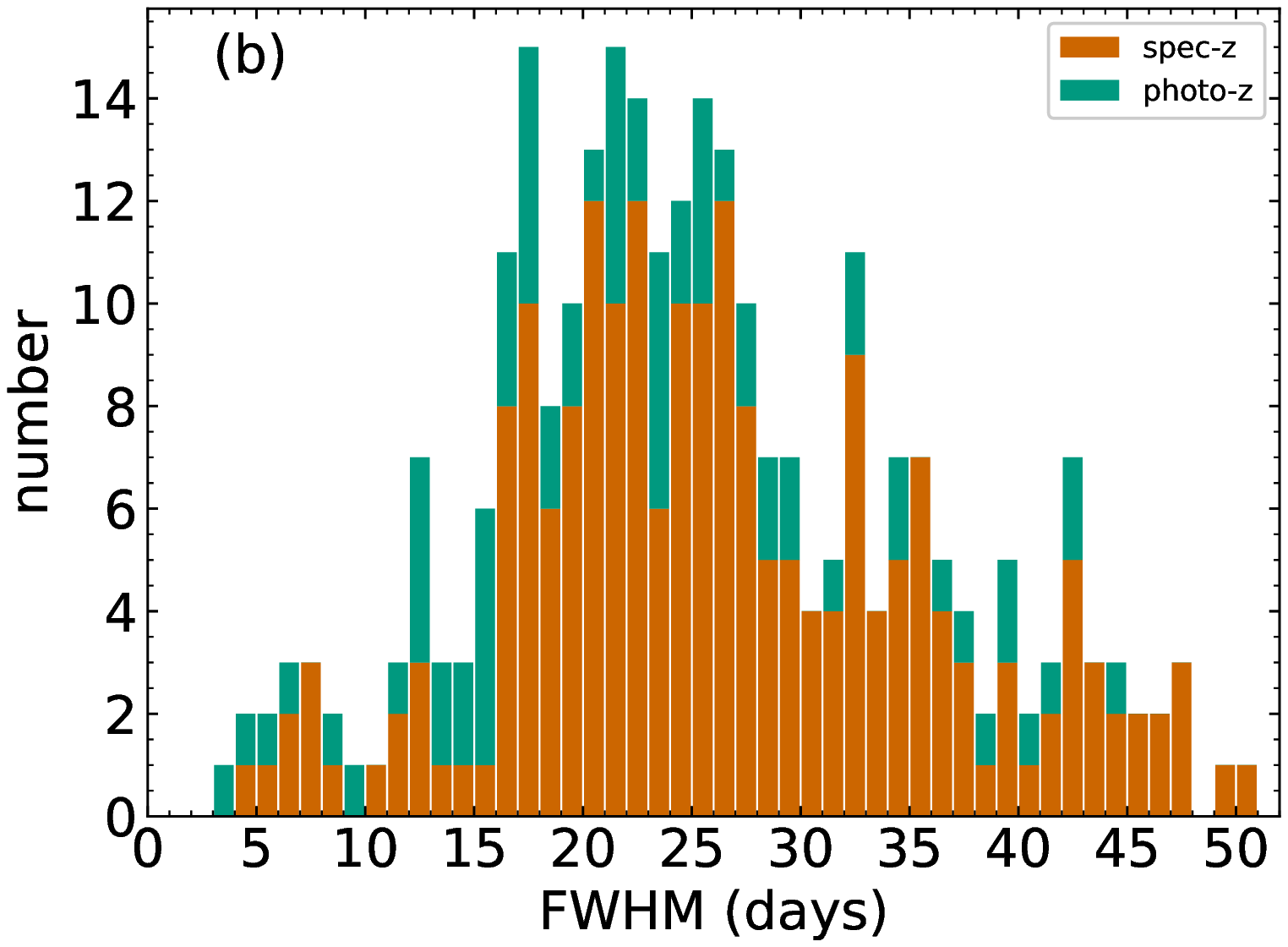}
 \end{center}
 \caption{The distributions of rest-frame FWHMs in the $i$-band for transients classified as genuine SNe Ia (a) and other transients (b) from the HSC-SSP Transient Survey.}
 \label{fig:FWHM}
\end{figure}

To identify rapidly evolving transients from the 1824 objects, the timescale of each sample has to be quantified.
\blue{
\citet{PS1} classified their rapidly evolving transients based on the rise and decline rates from the peak in the observed frame. \citet{DES} performed a Gaussian fitting to the light curve in the observed frame and selected the rapidly evolving transients according to the full width half maximum (FWHM) of their best fit Gaussian function.}
In this paper, we perform a Gaussian fitting to the light curves in the rest frame by using the spectroscopic redshifts (if available) or photometric redshifts of host galaxies taken from COSMOS2015 catalog \citep{laigle16} and HSC photometric redshift catalog \citep{tanaka_photoz}. 
Then we select rapidly evolving transients based on the FWHM of the best fit Gaussian function for each object in the rest frame. Since the shapes of the light curves of rapidly evolving transients are still uncertain, we adopt the Gaussian function not to rely on specific assumption on the light curve shapes by following  \citet{DES}. 
In Figure $\ref{fig:gaussianedsample}$, we present examples of the Gaussian fitting for two objects: one is classified as genuine SN Ia by template fitting and the other is not. As shown in Figure \ref{fig:gaussianedsample}, the Gaussian function in flux gives a reasonable fitting, which is adequate to systematically quantify the light curve width. Figure \ref{fig:FWHM} shows the distribution of the measured FWHMs in the $i$-band.

\begin{figure*}[htbp]
 \begin{center}
  \includegraphics[width=59mm]{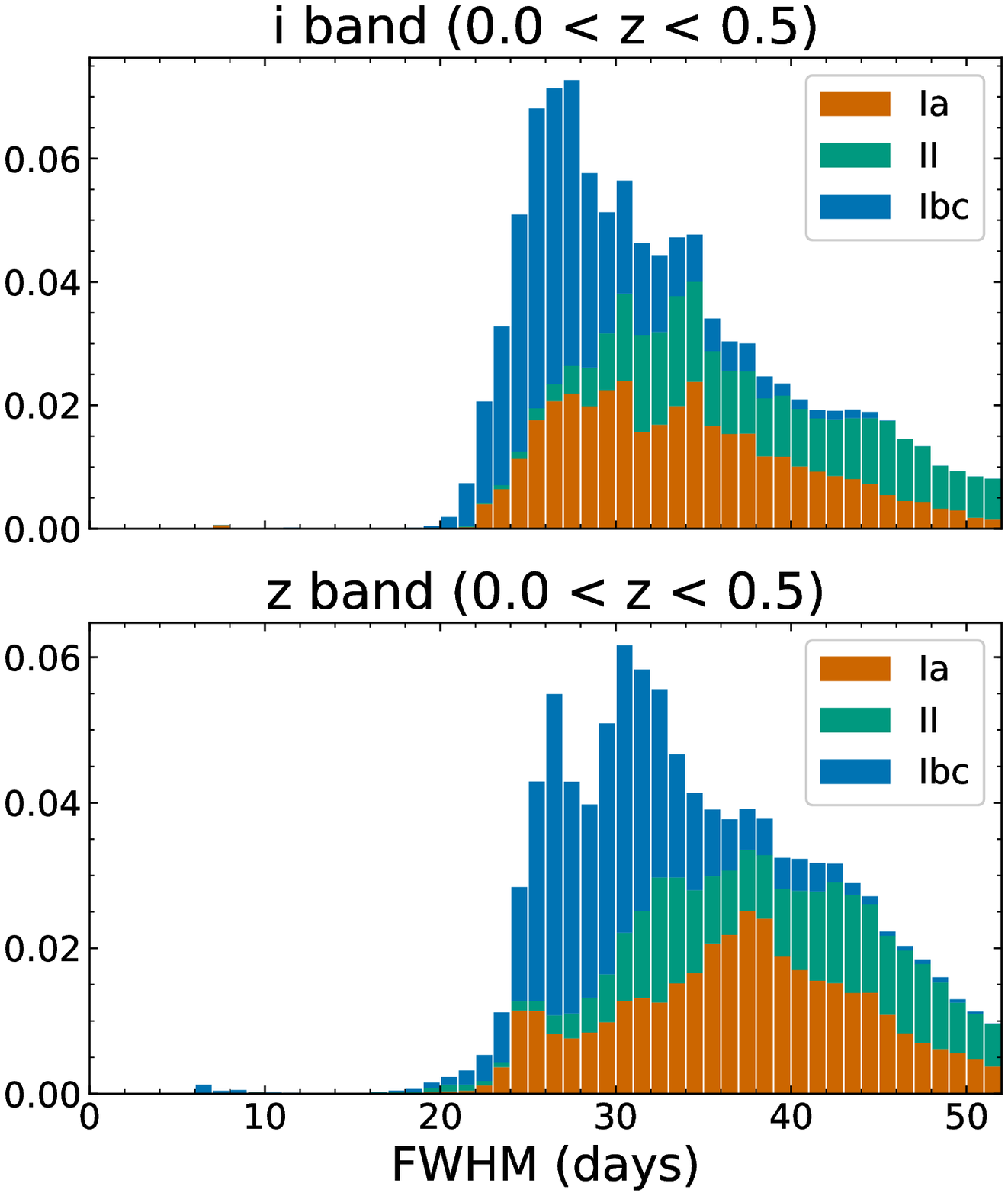}
  \includegraphics[width=59mm]{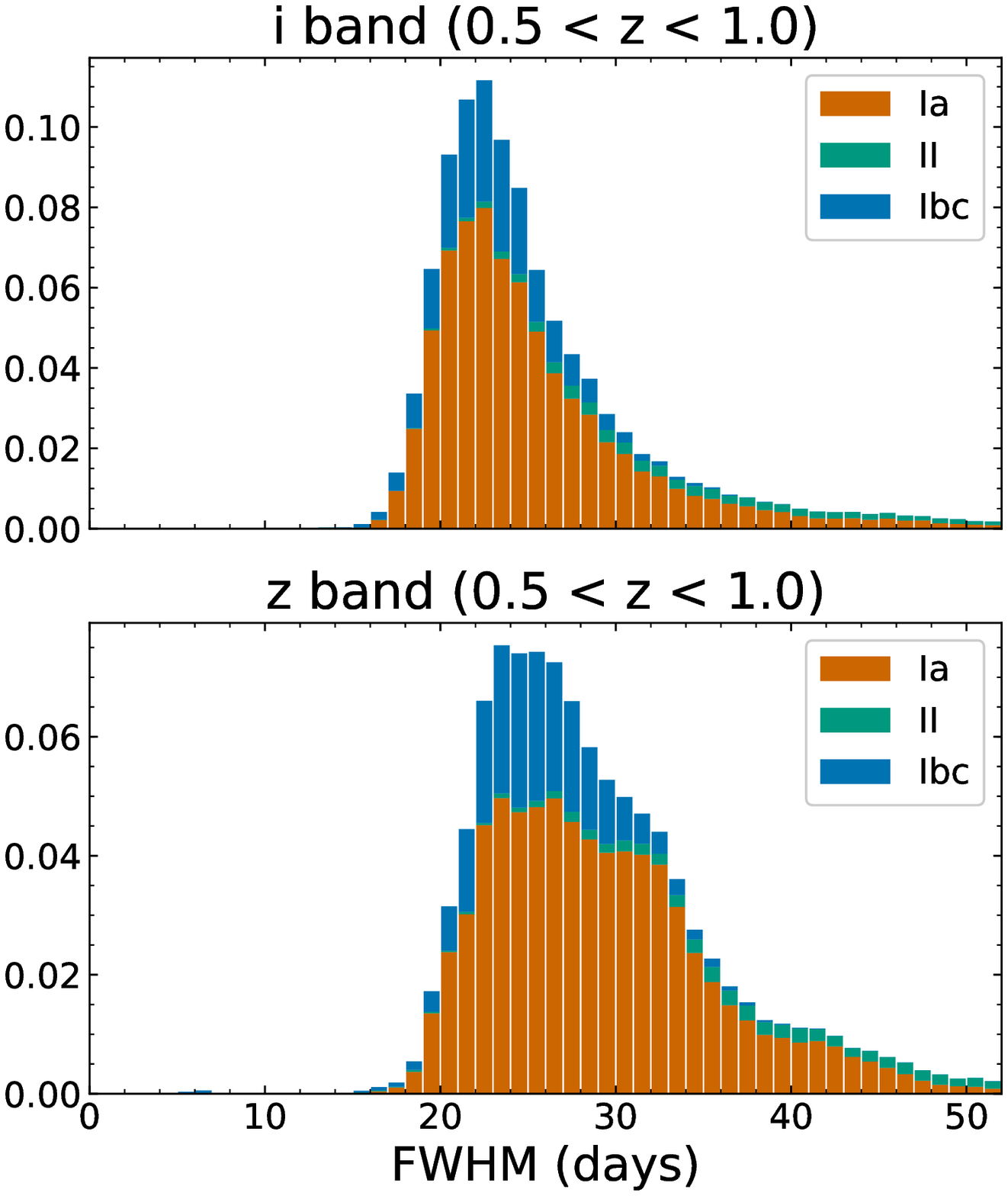}
  \includegraphics[width=59mm]{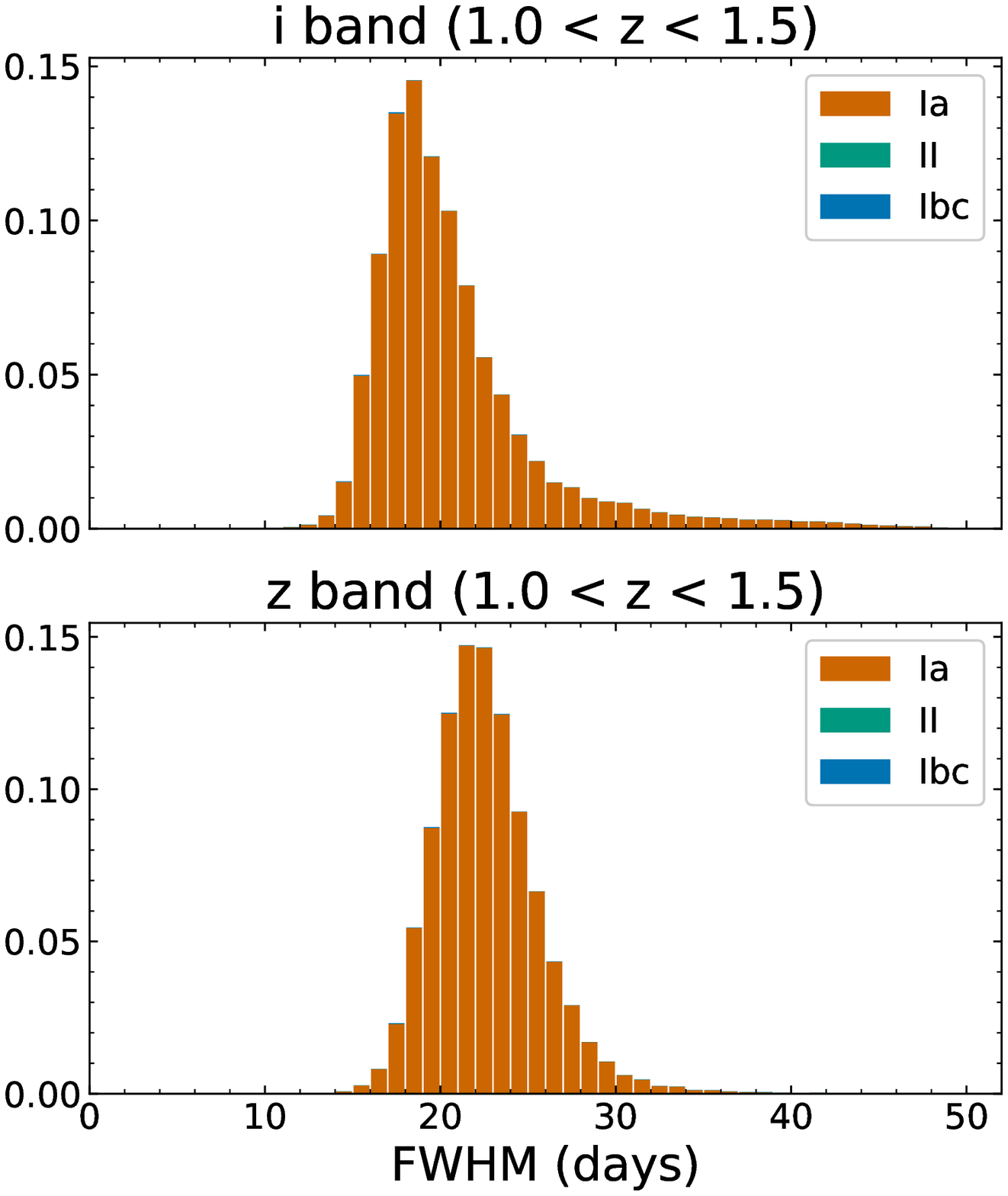}
 \end{center}
 \caption{The distributions of the rest-frame FWHMs in $i$-band (upper panels) and $z$-band (lower panels) for simulated SNe Ia, II and Ibc. Three different panels for each type of SN shows results of different redshift ranges: 0.0 $<$ z $\le$ 0.5 (left), 0.5 $<$ z $\le$ 1.0 (middle), and 1.0 $<$ z (right). The event rate of each type of SNe \citep{sn_fraction} is taken into account for the relative number of simulated SNe }. 
 \label{fig:FWHM_sim}
\end{figure*}

In order to define the criteria to select rapidly evolving transients, 
we generated light curves of normal SNe Ia, II and Ibc using available templates (SNe Ia: SALT2 template from \citealt{guy07}, SNe II and Ibc:  the SNANA packages from \citealt{kessler09}). Then we derived FWHMs of the simulated light curves by the Gaussian function fitting using sncosmo package \citep{barbary16}.
For the simulations, we first assume observing schedule and sensitivity of the HSC-SSP Transient Survey. Then we generate mock SNe by assuming the peak absolute magnitudes of each type of SNe as follows: $M_{\rm b,peak} = -19.05 \ - 0.141 \ x1  \ + 3.101 \ c$, where $x1 = 0.945 \pm 0.257$ and $c = -0.043 \pm 0.053$ for SNe Ia  \citep{Ia_params, Ia_rel, Betoule14}, $M_{\rm b,peak} \sim$ $-$16.75$\pm$0.98 for SNe II \citep{sn_mag}, and $M_{\rm b,peak} \sim$ $-$17.5$\pm$1.12 for SNe Ibc \citep{sn_mag}.
For the SNe Ia template, $x1$ and $c$ are stretch and color parameters, respectively.
\blue{
We assume zero extinction for the simulations of SN II and Ibc. This assumption does not affect the FWHM distribution, and only affects the maximum redshift at which SNe can be detected.
}
In Figure $\ref{fig:FWHM_sim}$, we show the distribution of the derived FWHMs for each type of simulated SNe in three different redshift ranges. We note that, since SNe II and Ibc are fainter than SNe Ia, a large fraction of simulated SNe II and SNe Ibc at z $\ge$ 0.6 do not satisfy our detection criteria (see below for more details).

\begin{figure*}[tbp]
 \begin{center}
  \includegraphics[width=88mm]{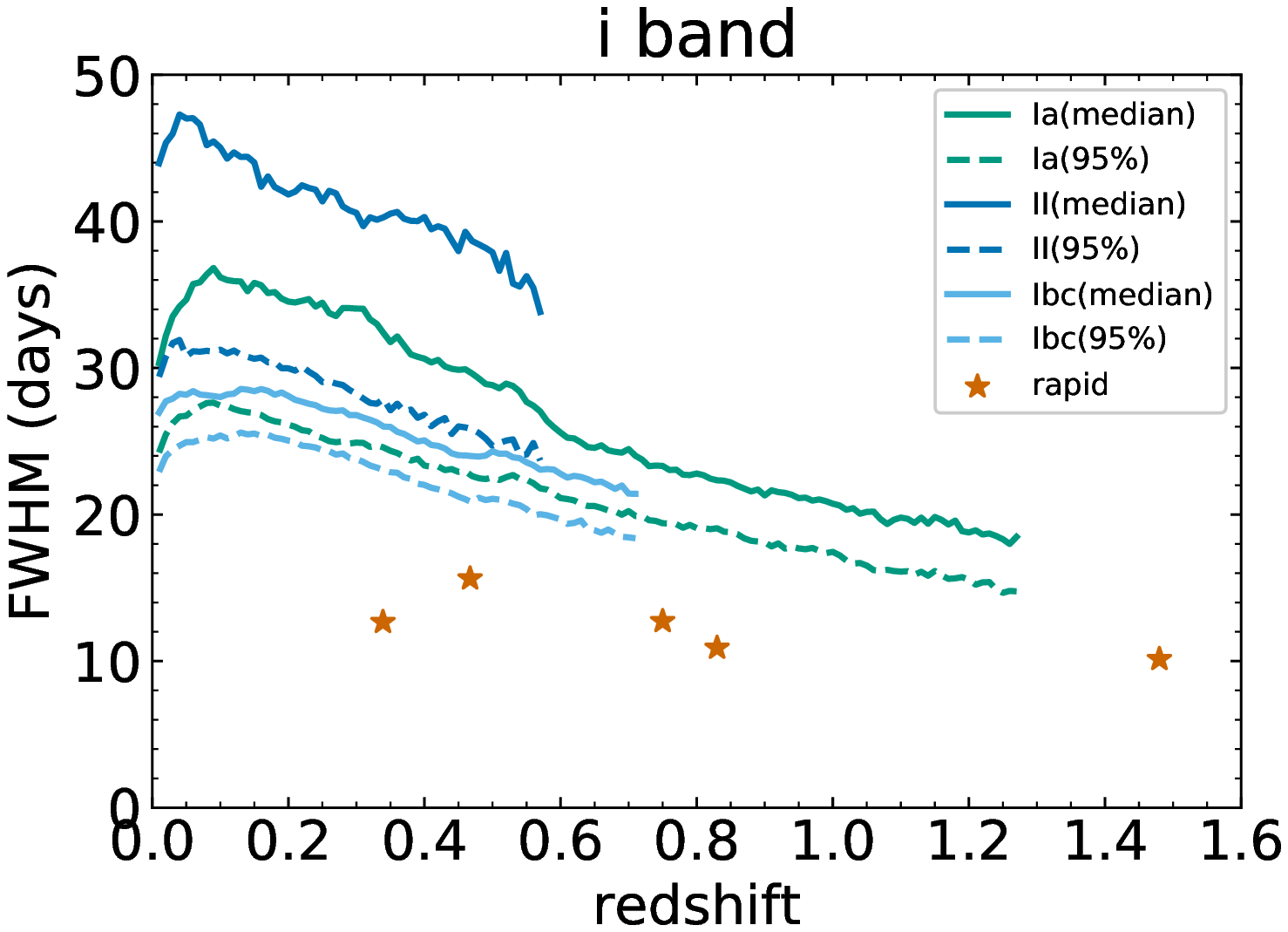}
  \includegraphics[width=88mm]{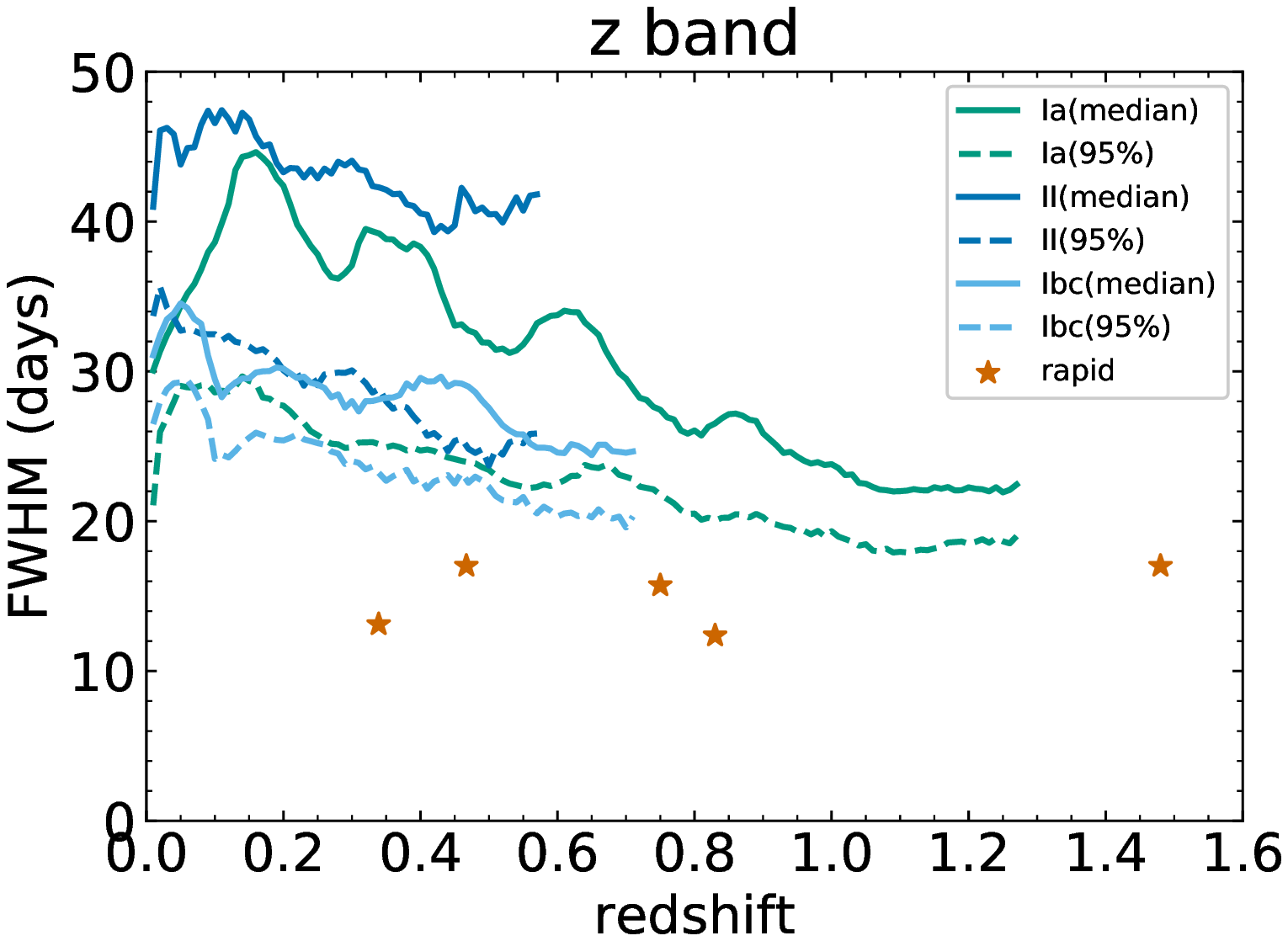}
 \end{center}
 \caption{The rest-frame FWHMs of the simulated
 SNe (lines) and those of the rapidly evolving transients found in the HSC-SSP Transient Survey (stars) in $i$-band (left) and $z$-band (right).
 The solid and dashed lines represent the distribution of FWHMs for each type of simulated SNe: the solid line is the median, and the dashed line is defined so that 95$\%$ of simulated SNe have the larger FWHMs than this line.} 
 \label{fig:FWHM_min}
\end{figure*}

Figure $\ref{fig:FWHM_min}$ presents the redshift dependence of the FWHMs of the simulated SNe. Note that the cutoff redshift for the simulated SNe is set by the peak magnitude of each type of SNe and the sensitivity of the HSC-SSP Transient Survey.
It is worth noting that the FWHMs of the light curves in the rest frame in a certain observed filter depend on the redshift. This is primarily due to the fact that, at higher redshifts, we observe shorter wavelengths in the rest frame, where the intrinsic light curves tend to be narrower. Another factor is that the number and significance of the detection is lower at high redshifts. Then the light curves tend to be sampled only around the maximum. When the slower decline after the peak is missed, the FWHM tends to be shorter for such light curves. Therefore, we decided to impose redshift-dependent criteria for the FWHMs.

Through the measurements and simulations described above, we define the criteria to select rapidly evolving transients from the 1824 objects as follows:\\
1 : More than 5 detections in total beyond 5$\sigma$ in the 4 bands ($g$, $r$, $i$, and $z$).\\
2. The light curves have data in both pre- and post-maximum, both in the $i$- and $z$-bands.\\
3. Redshift of a host galaxy is determined by spectroscopy or photometric redshift with an error smaller than 10$\%$ is available.\\
4. The best-fit reduced $\chi^2$ of the template fitting (SNe Ia, II, and Ibc) is greater than 2.0.\\
5. The FWHMs in the rest frame satisfy the following conditions.\\
5a. 0 $<$ $z$ $\le$ 0.5: FWHM in the $i$-band is less than 23 days and FWHM in the $z$-band is less than 24 days.\\
5b. 0.5 $<$ $z$ $\le$ 1.0 :  FWHM in the $i$-band is less than 18 days and FWHM in the $z$-band is less than 20 days.\\
5c. 1.0 $<$ $z$ : FWHM in the $i$-band is less than 15 days and FWHM in the $z$-band is less than 18 days.\\
We set the first, second, and third criteria to accurately measure their timescales in the rest frame. 
The fourth criterion is set to remove normal types of SNe.
For the fifth criteria, we only use the $i$- and $z$-bands
because the HSC-SSP Transient Survey has more epochs in the $i$- and $z$-bands than in the $g$- and $r$-bands, and thus, the FWHMs of the $i$- and $z$-bands are more reliable.

\blue{
Note that \citet{arcavi_rapid} also searched for transients with short timescales. Their aim was to find out the transients which showed a rapid rise ($\sim$10 days) to a luminous peak ($\sim 5 \times 10^{43} \rm \ erg \ s^{-1}$).  
Our criteria are tuned to select the transients with a very short timescale regardless of a peak luminosity, and not designed to identify the object like transients reported by \citet{arcavi_rapid}.
}

\begin{figure*}[bp]
 \begin{center}
   \includegraphics[width=57mm]{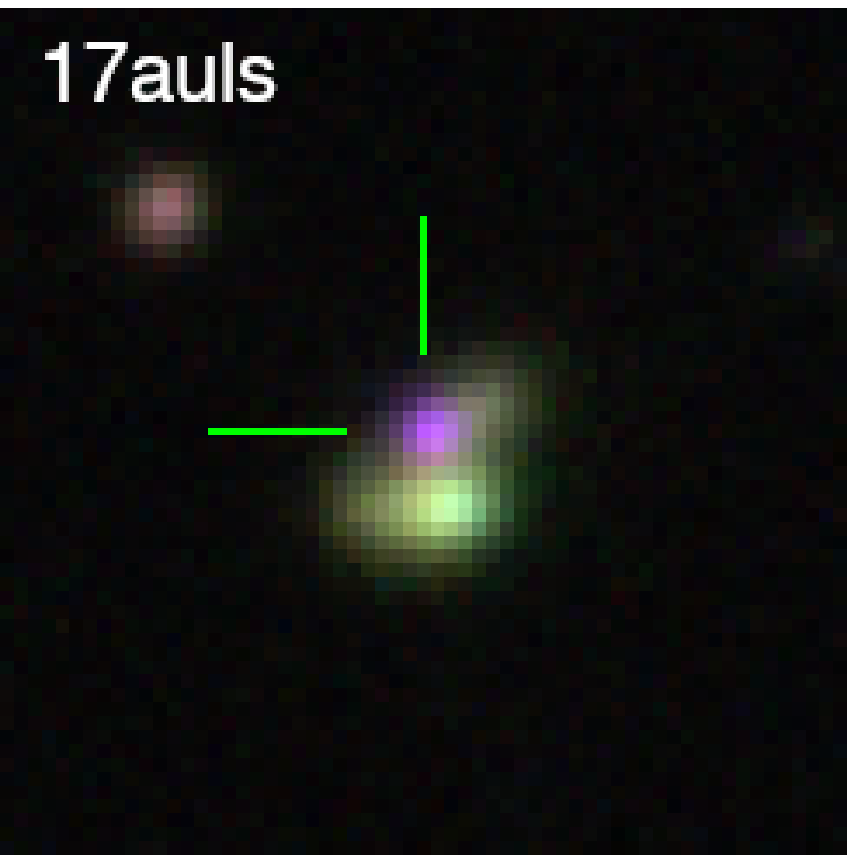}
   \includegraphics[width=57mm]{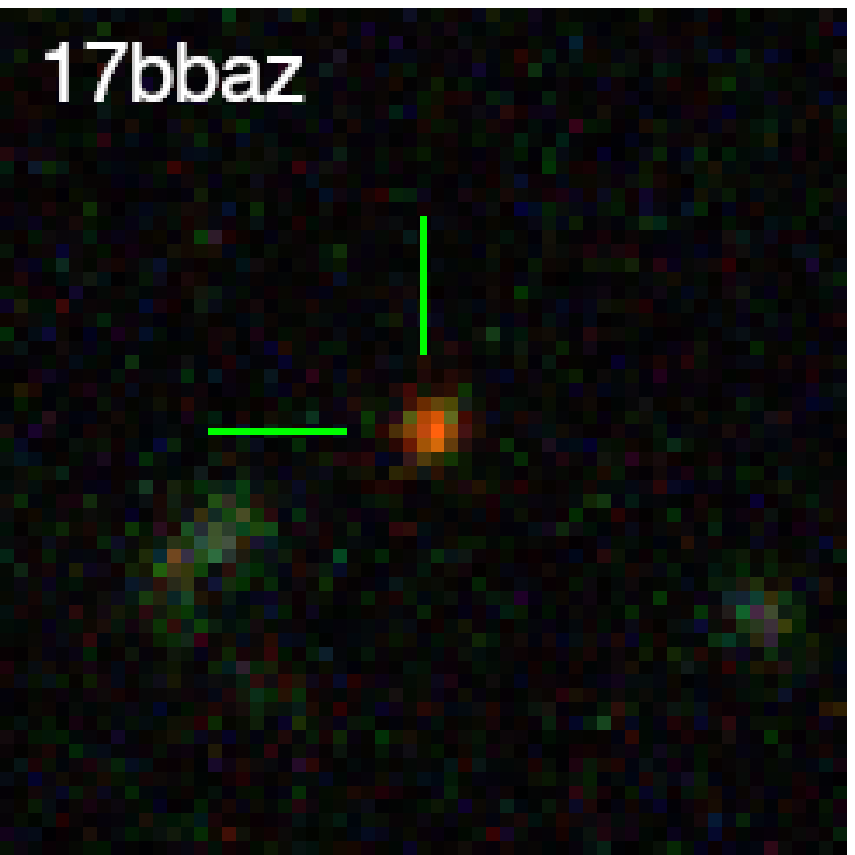}
   \includegraphics[width=57mm]{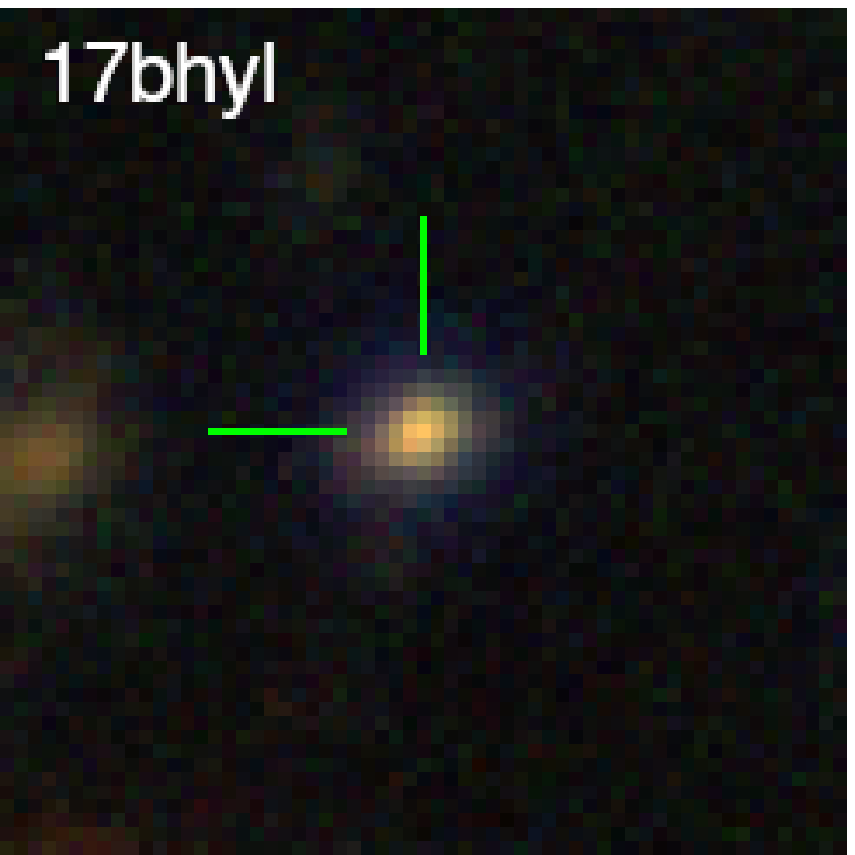}
   \includegraphics[width=57mm]{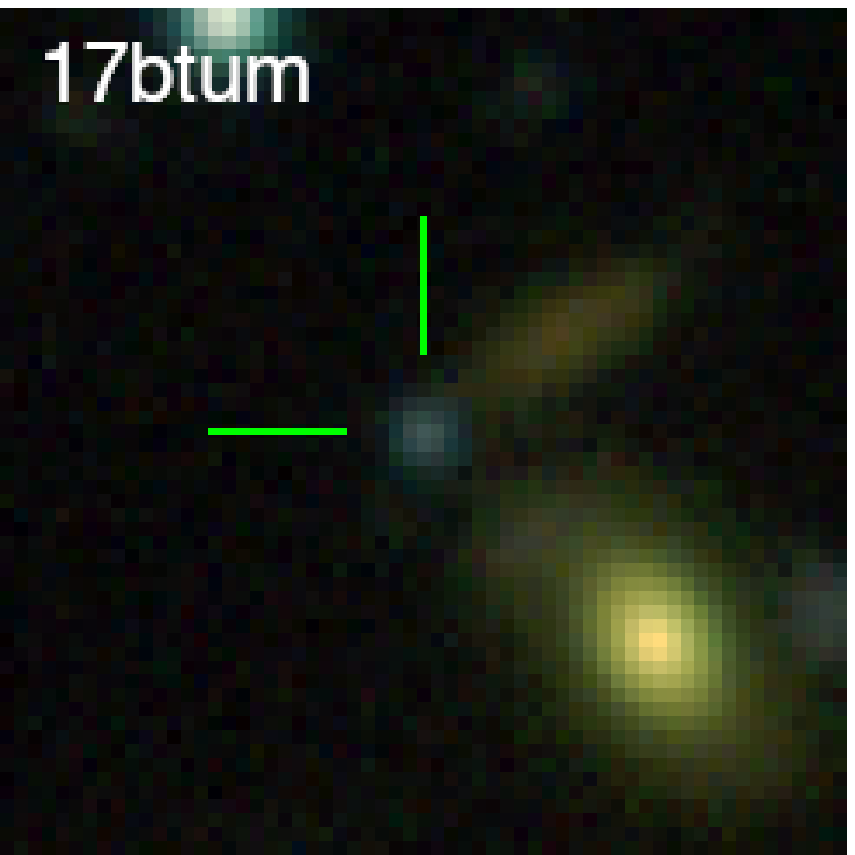}
   \includegraphics[width=57mm]{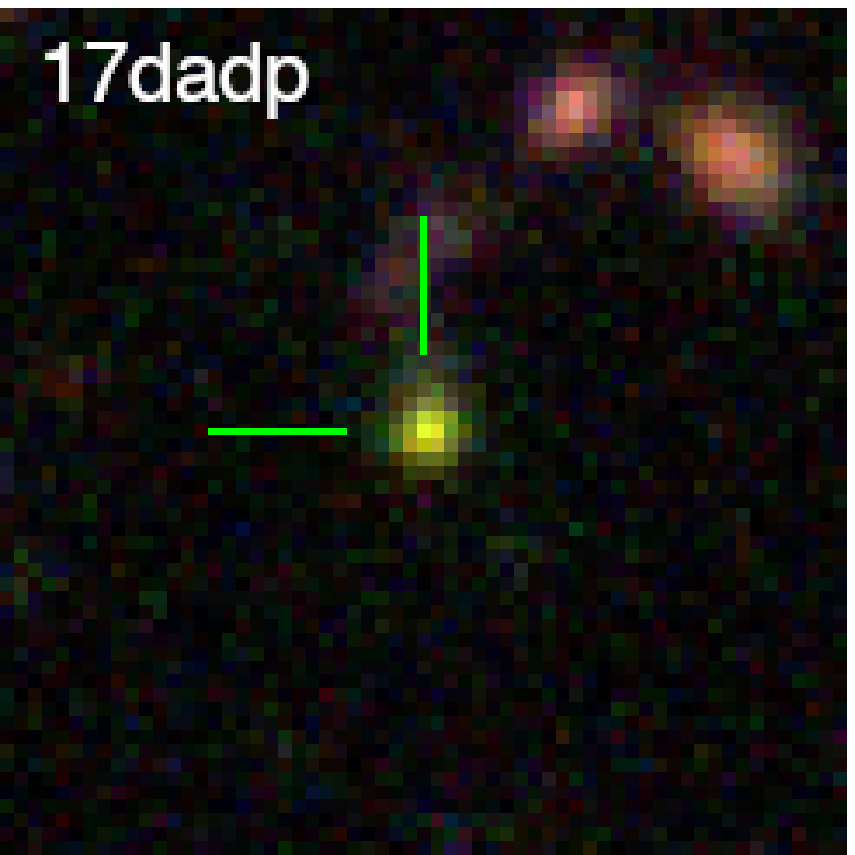}
  \end{center}
 \caption{Three-color images of the five rapidly evolving transients from the HSC-SSP Transient Survey. Red, green and blue colors represent the $i$-, $r$- and $g$-band images, respectively. Each image has a size of 10.2 arcsec $\times$ 10.2 arcsec. North is up and east is to the left.}
 \label{fig:RETwithhost}
\end{figure*}

By applying these criteria, we identified 5 objects as the rapidly evolving transients from the HSC-SSP Transient Survey (Figure $\ref{fig:RETwithhost}$). 
In the following sections, we present the properties of these transients.

\section{sample overview}
\label{sec:samples}

In this section, we give an overview of the rapidly evolving transients identified with the HSC-SSP Transient Survey.
In Table $\ref{tab:RET}$, we summarize basic information of our samples.
For two objects among our samples (HSC17auls and HSC17bhyl), redshifts are obtained through spectroscopic observations of their host galaxies.
\blue{For the other three transients, the redshifts are estimated by multi-color photometric data: COSMOS2015 catalog \citep{laigle16} for HSC17btum and HSC-SSP photo-z catalog \citep{HSC_photoz} for HSC17bbaz and HSC17dadp. Photometric redshifts of HSC-SSP are estimated by different methods: MLZ \citep{MLZ}, EPHOR \footnote{\url{https://hsc-release.mtk.nao.ac.jp/doc/index.php/photometric-redshifts/}} and Franken-z (Speagle et al. in prep). 
For the host galaxies of HSC17bbaz and HSC17dadp, all the methods give the consistent estimates within the error. We use the values from Mizuki in Table $\ref{tab:RET}$ and following discussion.}

Note that our criteria for FWHMs are quite uncertain at redshifts beyond 1.0 (Figure $\ref{fig:FWHM_min}$) due to the poor understanding of spectral energy distribution of normal SNe at rest frame UV wavelengths. 
\blue{
In fact, the redshift of the host galaxy for HSC17bbaz is $z = 1.48$.
By assuming this redshift, 
the offset from the host galaxy is rather large ($25.69$ kpc, see Section \ref{sec:galaxies}).
Therefore, we cannot fully exclude the possibility that the real host galaxy of HSC17bbaz is located near the position of the transient but it is not detected by the HSC-SSP survey. For example, if the real host galaxy were located at $z = 1$, the FWHM would be 20 and 21 days in the $i$ and $z$ band, and the peak magnitude would be $-18.95$ mag in the $i$ band. Then, the properties of HSC17bbaz are consistent with SNe Ia. For this case, since the limiting magnitude of the reference image of HSP-SSP Transient Survey is $\sim$ 27 mag, the host galaxy must be fainter than -17 mag in absolute magnitude, which might also be possible.
By these reasons, we regard HSC17bbaz ($z=1.480$) as a marginal candidate of a rapidly evolving transient. 
}

\begin{figure*}[htbp]
 \begin{center}
   \includegraphics[width=82mm]{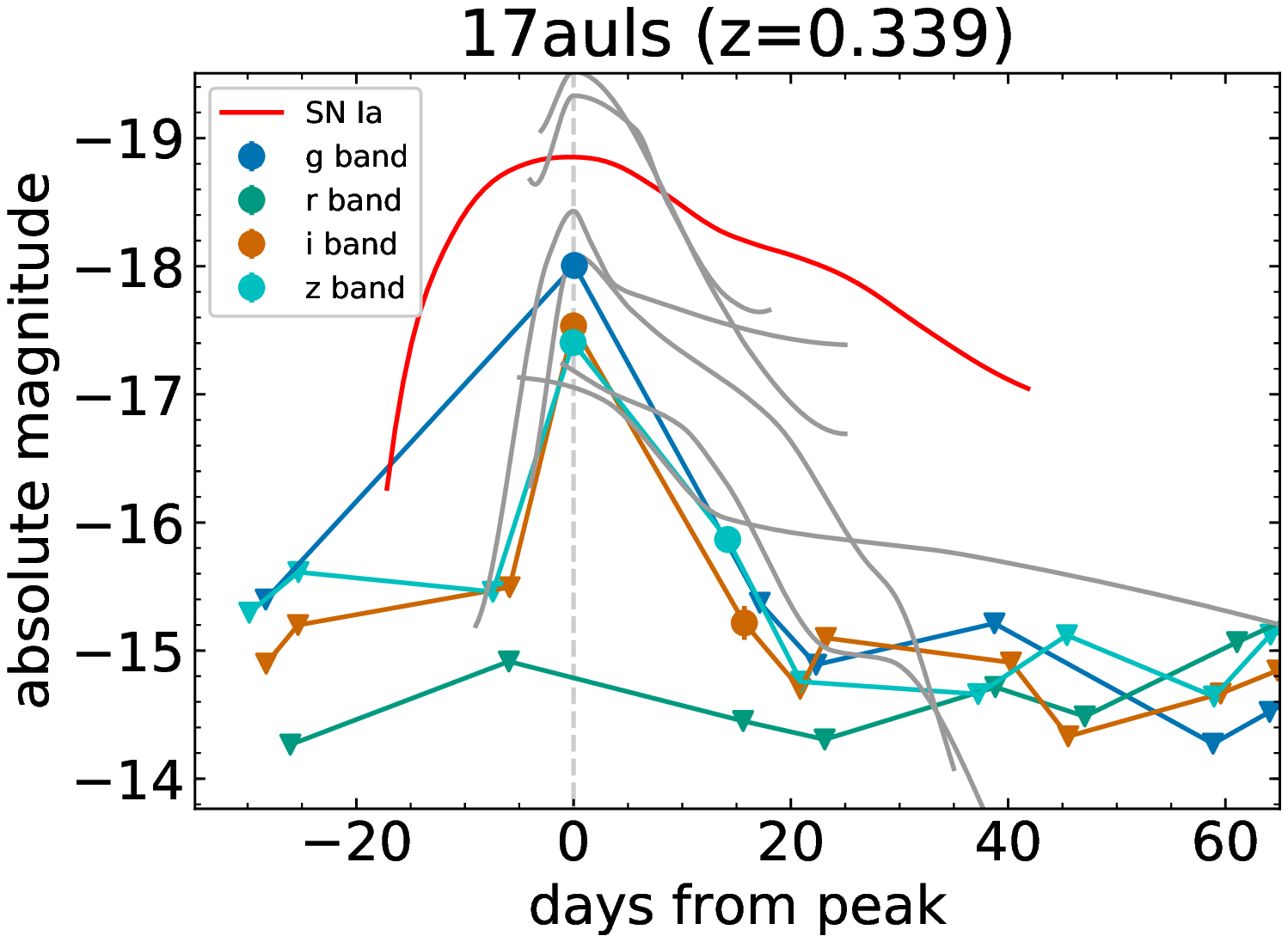}
   \includegraphics[width=82mm]{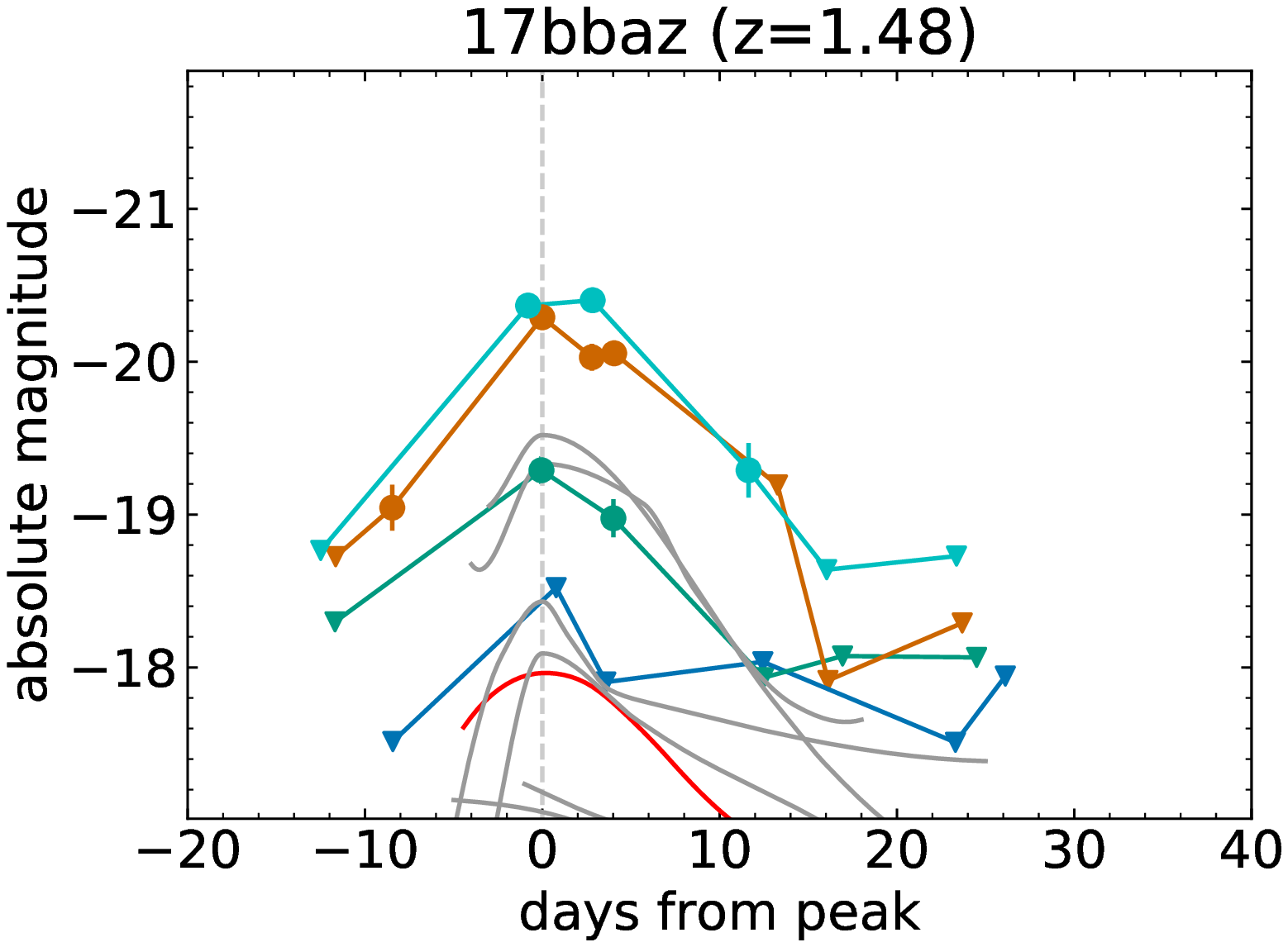}
   \includegraphics[width=82mm]{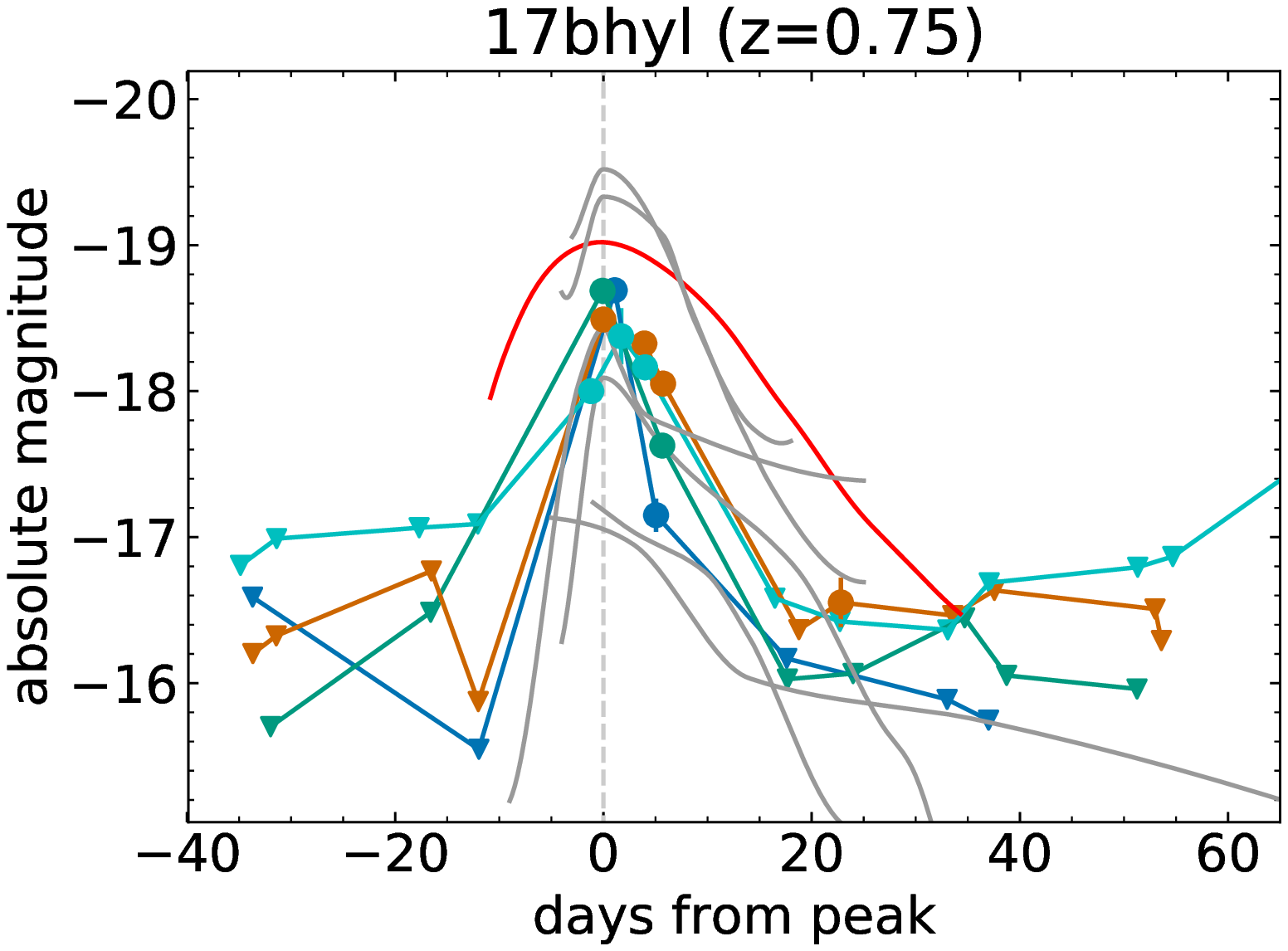}
   \includegraphics[width=82mm]{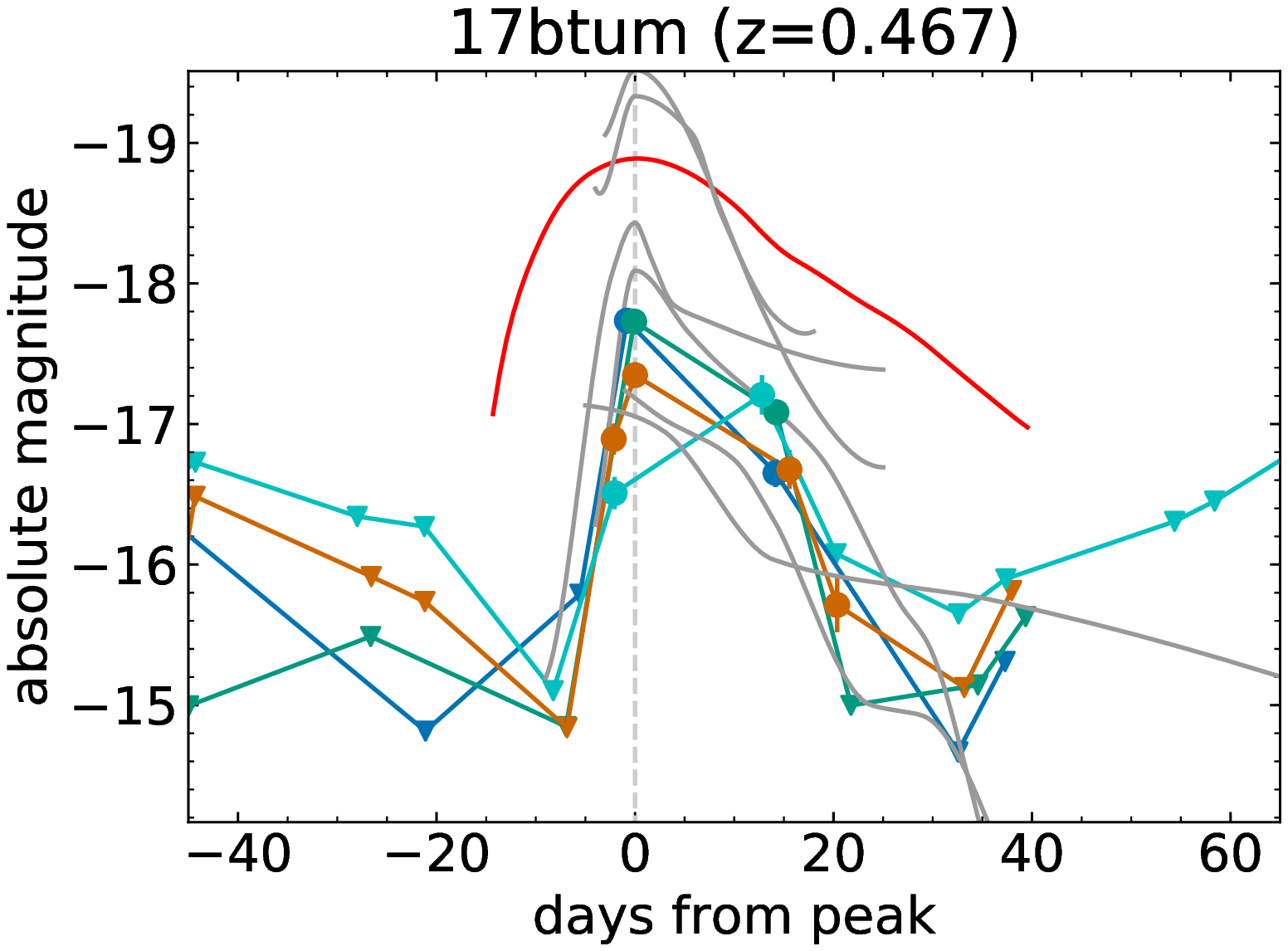}
   \includegraphics[width=82mm]{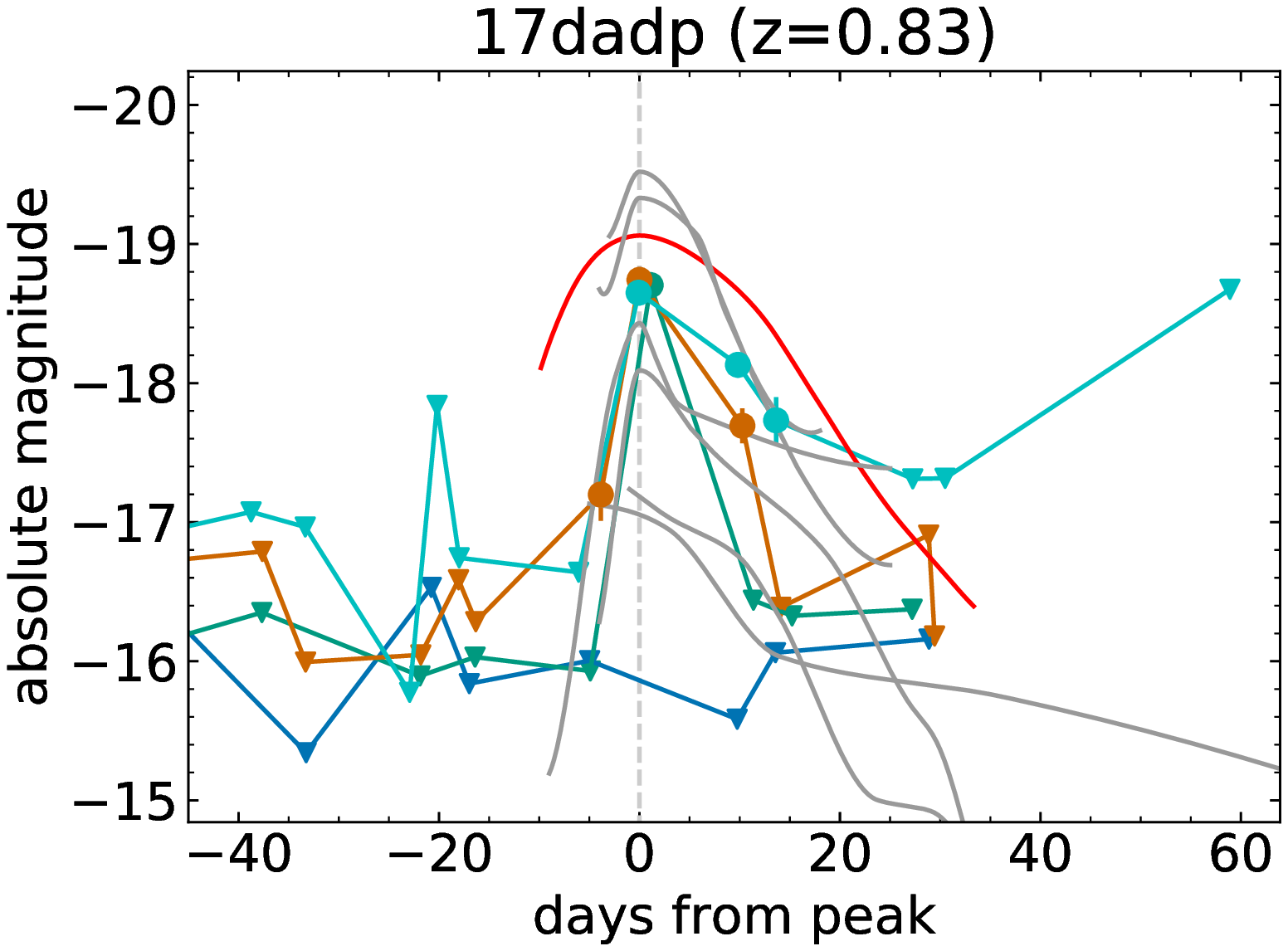}
 \end{center}
 \caption{Absolute-magnitude light curves of our samples with respect to the peak epoch in the observed $i$-band. 
 The absolute magnitudes are given with a rough $K$-correction, $M = m - \mu + 2.5 \log(1+z)$.
 The epoch is given in the restframe of the transients. The circles represent the detections while the triangles represent the 3 $\sigma$ upper limits in the observed $g$-, $r$-, $i$- and $z$-bands. The red solid lines show SN Ia light curves in the observed $i$-band generated by the sncosmo package using the SALT2 model ($E(B-V)$ = 0.0, $x1$ = 0.945, $c$ = $-$0.043,  \citealt{Ia_params}) at the same redshift of each object.
 The gray lines show the light curves of 6 rapidly evolving transients in the observed $i$-band from PS1 \citep{PS1}.}
 \label{fig:maglightcurve}
\end{figure*}

We present the absolute-magnitude light curves of our samples in the rest frame (Figure $\ref{fig:maglightcurve}$). The light curves are shown with respect to the $i$-band peak epoch.
\greenn{The log of the photometric observations for our samples in the observed frame is listed in Table \ref{tab:obs_log}.}
For the absolute magnitude in Figure \ref{fig:maglightcurve}, a simple $K$-correction is applied as $M = m - \mu + 2.5 \log(1+z)$, \green{by considering only the term of the cosmic expansion from the full $K$-correction \citep{Hogg02}. Here $M$ is the absolute magnitude and $m$ is the observed magnitude.}
The red solid line represents a light curve of SN Ia simulated at the same redshift by using the SALT2 model with the mean value of parameters ($E(B-V)$ = 0.0, $x1$ = 0.945, $c$ = $-$0.043; \citealt{Ia_params}). The gray lines show the $i$-band light curves of the rapidly evolving transients found in PS1 \citep{PS1}. All of our rapidly evolving transients show timescales shorter than SNe Ia, which confirms the validity of our selection criteria.
And the timescales of our samples are similar to those of the rapidly evolving transients found by PS1. 

\begin{figure*}[htbp]
 \begin{center}
   \includegraphics[width=180mm]{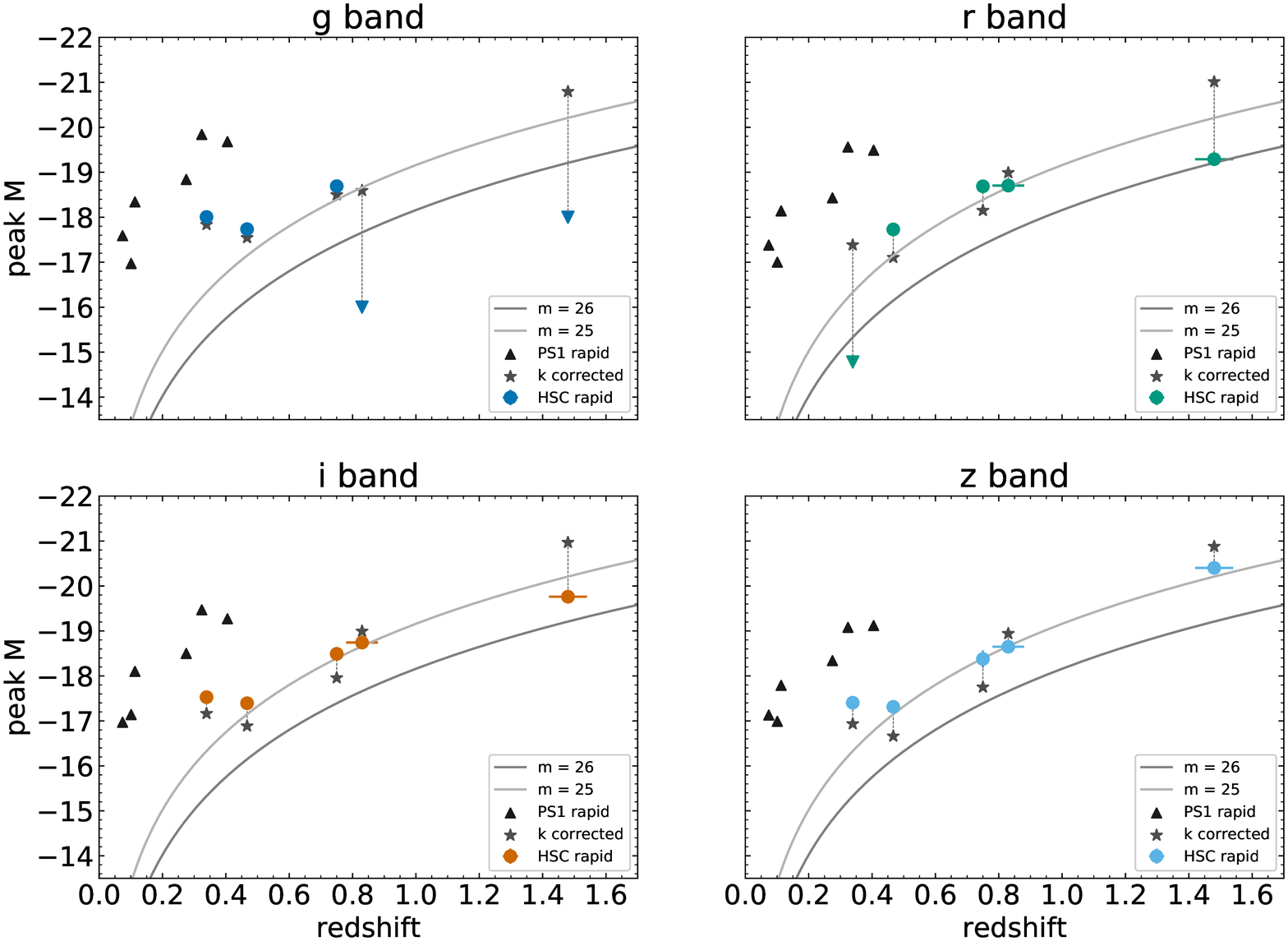}
 \end{center}
 \caption{Peak absolute magnitudes as a function of redshift for our samples in each observed bands ($g, r, i$ and $z$). Circles represent the observed peak absolute magnitude after a rough $K$-correction ($M = m - \mu + 2.5 \log(1+z)$).
 Stars represent the absolute magnitude $K$-corrected to the {\it rest} frame based on the blackbody fitting (see Section $\ref{sec:blackbody}$). Triangles represent the $K$-corrected magnitude of the rapidly evolving transients from PS1 in the {\it rest} frame. The solid lines correspond to the observed magnitude limits of 25 mag and 26 mag.}
 \label{fig:peakvsz}
\end{figure*}

In Figure $\ref{fig:peakvsz}$, we summarize the peak absolute magnitudes as a function of redshift in each observed band. The redshifts and peak absolute magnitudes have a large variety with the range of 0.339 $\le z \le$ 1.480 and $-$19.76 $\le$ M$_i$ $\le$ $-$17.53. Compared to the samples from  PS1 \citep{PS1} and Dark Energy Survey \citep{DES}, the redshifts of our samples tend to be higher while the range of peak magnitude is similar.

\startlongtable
\begin{deluxetable*}{c|rrlc}
    \tablecaption{\green{Basic observational information of the rapidly evolving transients from HSC-SSP Transient Survey}
    \label{tab:RET}}
    \tablehead{
      \colhead{Name} & \colhead{R.A.} & \colhead{Dec.} & \colhead{Redshift} & \colhead{Redshift source}\\
     \colhead{} & \colhead{J2000.0} & \colhead{J2000.0} & \colhead{} & \colhead{}
    } 
    \startdata
    HSC17auls & 09:57:55.13 & +02:25:08.10 & 0.339 & spec-z\\
    HSC17bbaz & 09:58:26.32 & +00:53:08.09 & 1.480$^{+0.063}_{-0.056}$ & photo-z \tablenotemark{a}\\
    HSC17bhyl & 10:01:22.21 & +02:01:53.37 & 0.750 & spec-z\\
    HSC17btum & 09:57:54.02 & +02:39:57.17 & 0.467$^{+0.010}_{-0.011}$ & photo-z \tablenotemark{b}\\
    HSC17dadp & 10:00:29.05 & +01:36:27.66 & 0.830$_{-0.047}^{+0.055}$ & photo-z \tablenotemark{a}
    \enddata
    \tablenotetext{a}{Photo-z is taken from the HSC-SSP Survey \citep{HSC_photoz}}
    \tablenotetext{b}{Photo-z is taken from the COSMOS2015 catalog \citep{laigle16}.}
\end{deluxetable*}

\section{photometric properties}
\label{sec:photometric}

\subsection{Timescale of the Light Curve}

\begin{figure*}[htbp]
 \begin{center}
   \includegraphics[width=180mm]{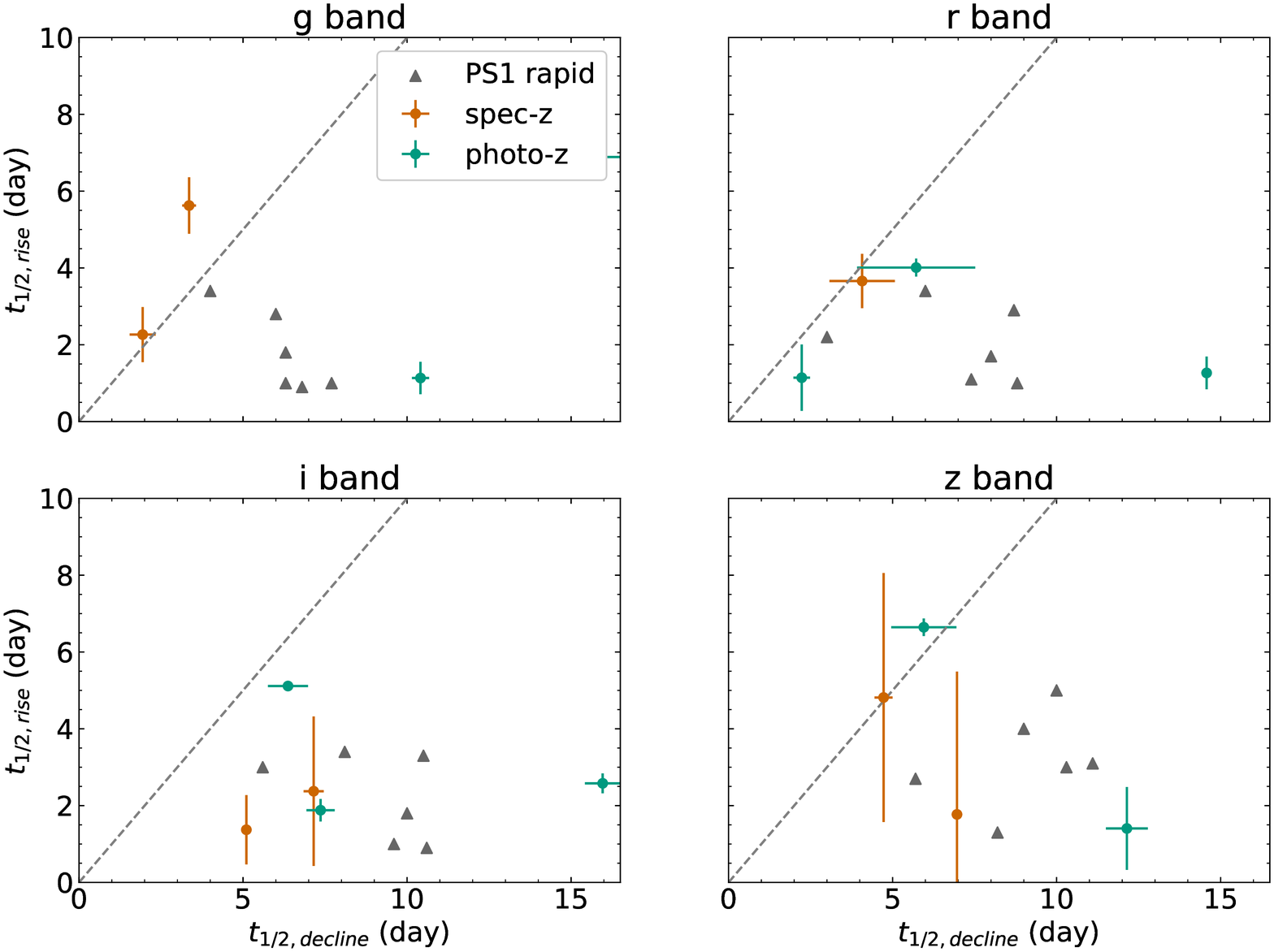}
  \end{center}
  \caption{Relation between the rise time  t$_{\rm 1/2,rise}$ and the declining time t$_{\rm 1/2,decline}$ in each observed band. Dashed lines represent the relation of $t_{\rm 1/2,rise}$ = $t_{\rm 1/2,decline}$. Gray triangles show the same for the rapidly evolving transients from PS1 \citep{PS1}}.
  \label{fig:risevsdecline}
\end{figure*}

In order to quantify the timescales of the light curves, we measure the rise and decline times for our samples. We define the rise ($t_{1/2, \rm rise}$) and decline times ($t_{1/2, \rm decline}$) as the times between the observed peak epoch and the epochs when the flux is half of the peak flux by linear interpolation in the magnitude.
The errors of timescale are also estimated by assuming linear time evolution. Thus, they may have large uncertainties depending on the timings of the observations.
Although we have an estimate of the peak epoch and light curve width from the Gaussian fitting, we use the observed peaks and measure the time with respect to it. This is because the peaks estimated by the Gaussian fitting can be biased to a later epoch when the light curve shows a slower declining tails.

In Figure $\ref{fig:risevsdecline}$, we present relations between the $t_{1/2, \rm rise}$ and $t_{1/2, \rm decline}$ in each observed band.
\green{We also summarized the measured $t_{1/2, \rm rise}$ and $t_{1/2, \rm decline}$ in Table \ref{tab:RET_lc}.}
In most of our samples, the rising time is shorter than the declining time.
Figure \ref{fig:risevsdecline} also includes the same timescales for the rapidly evolving transients from PS1 \citep{PS1}. Note that since the rest frame wavelength of each transient corresponding to the same observed band varies with redshift, {a comparison in a certain band is not straightforward.
Nevertheless, both the rise and decline times as well as the relation between them are similar to the results from PS1 \citep{PS1}.}

\subsection{Temperature and Luminosity}
\label{sec:blackbody}

\begin{figure*}[htbp]
 \begin{center}
   \includegraphics[width=55mm]{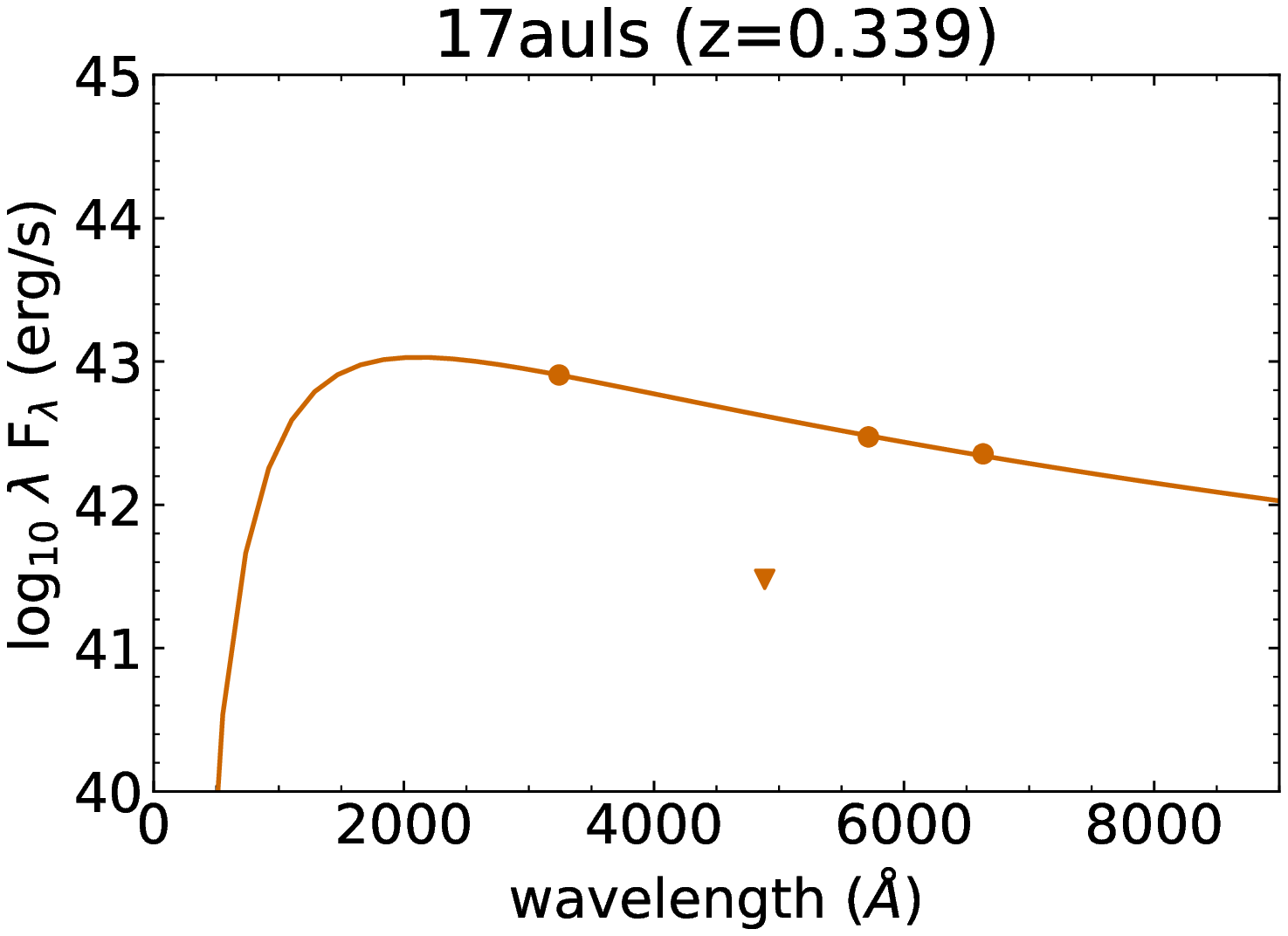}
   \includegraphics[width=55mm]{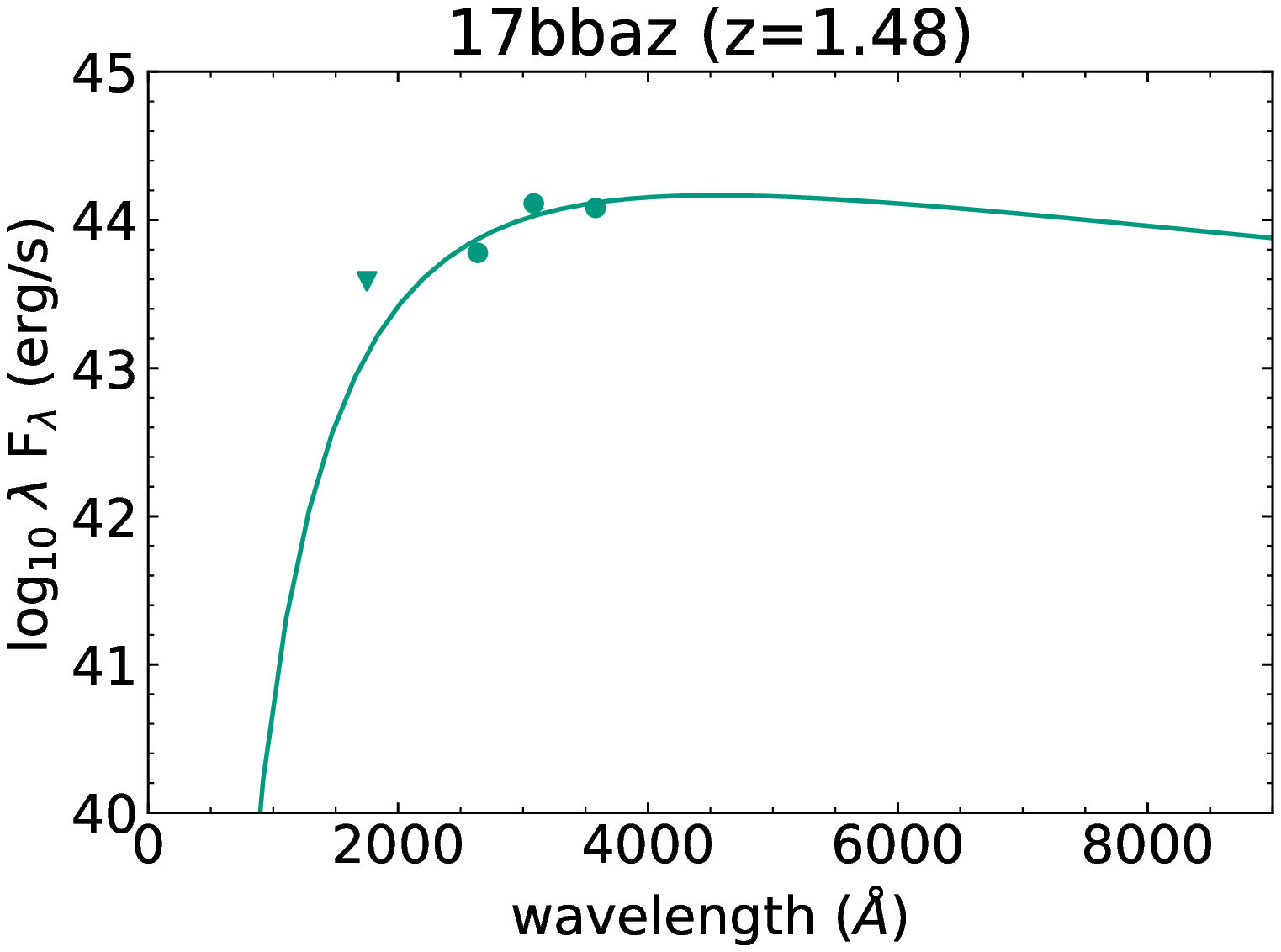}
   \includegraphics[width=55mm]{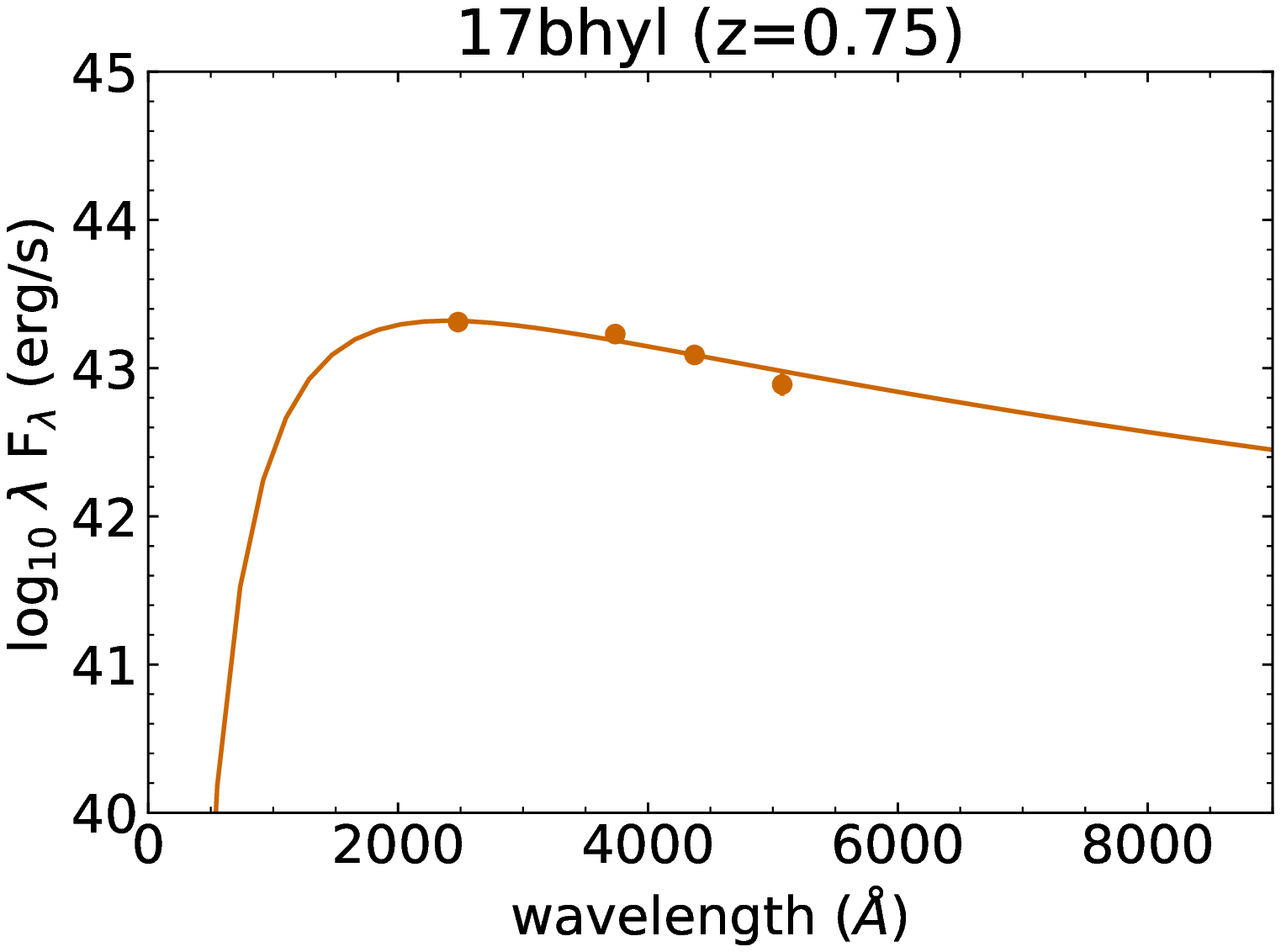}
   \includegraphics[width=55mm]{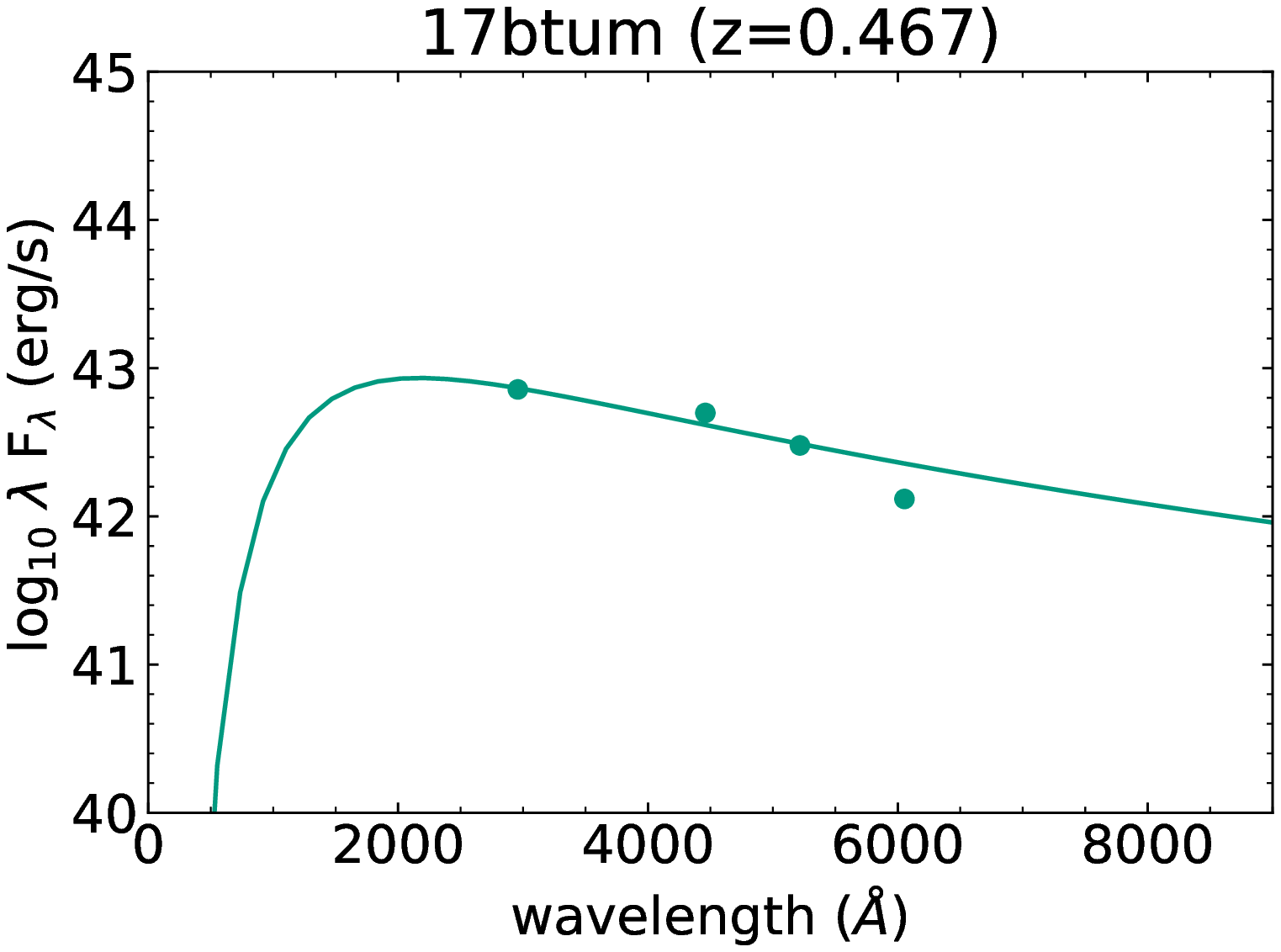}
   \includegraphics[width=55mm]{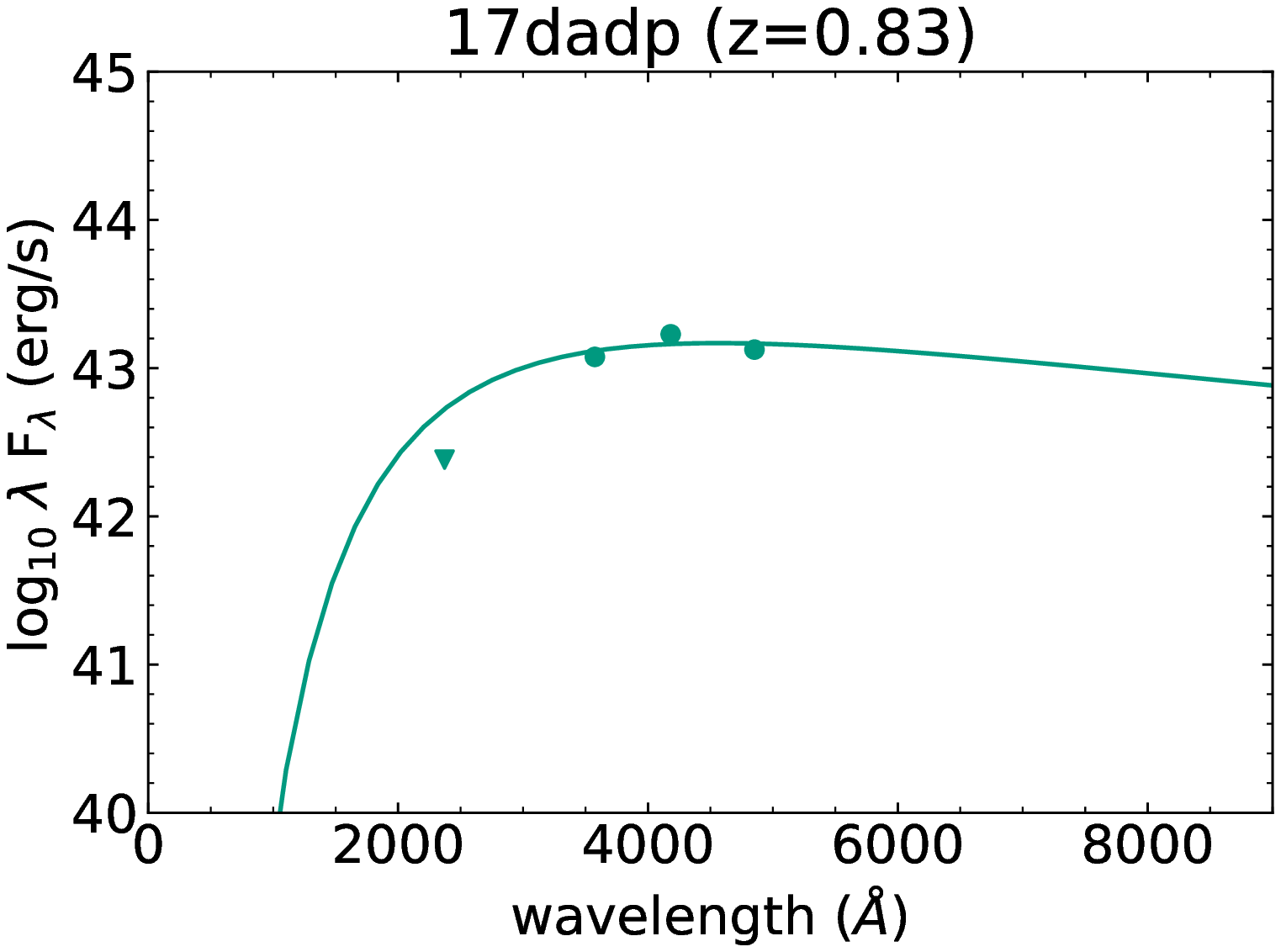}
 \end{center}
 \caption{Spectral energy distribution (in the rest frame) of the rapidly evolving transients from the HSC-SSP Transient Survey at the $i$-band peak epoch. Solid lines show the best-fit blackbody. The very faint flux of HSC17auls at 5000 \AA\ is an artifact, due to the lack of $r$-band data around the $i$-band peak epoch.}
 \label{fig:BBfit}
\end{figure*}

To characterize the spectral energy distributions of the rapidly evolving transients, 
we perform blackbody fitting to our samples and estimate temperatures and radii at their peak epochs.
Since we do not have all the four-band data in the same epoch, we use the peak epoch in the $i$-band and derive the flux in the other bands by linearly interpolating the magnitudes (see Figure \ref{fig:maglightcurve}). 
Then, we convert the magnitudes to flux densities in the rest frame, and perform blackbody fit in the rest frame.
\green{The estimated temperature at the peak for our samples is summarized in Table \ref{tab:RET_lc}} 
Figure $\ref{fig:BBfit}$ shows the best fit results of our samples.
We note that, HSC17auls lacks $r$-band observations around its $i$-band peak, so that the linear interpolation resulted in the very unrealistically peak magnitude of the $r$-band.

\begin{figure}[htbp]
 \begin{center}
  \includegraphics[width=85mm]{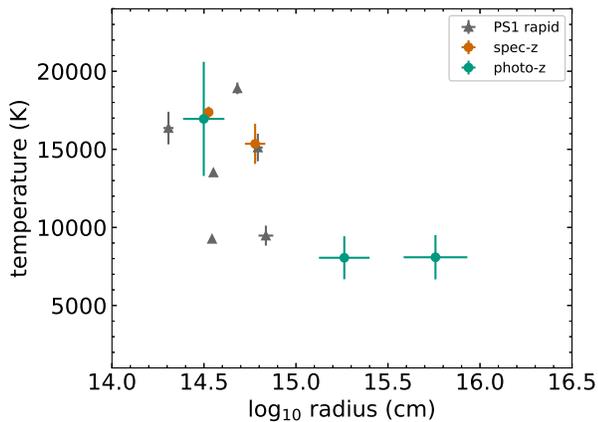}
 \end{center}
 \caption{Relation between the best-fit blackbody temperature and photospheric radius. Color points represent our samples and gray triangles represent the samples from PS1 \citep{PS1}}
 \label{fig:BBfit_TR}
\end{figure}

The derived temperatures range from 8,000 K to 18,000 K with a median of $\sim$14,000 K (see Figure $\ref{fig:BBfit_TR}$). The temperatures of HSC17bbaz and HSC17dadp, which are located at z $>$ 0.8, are lower than 10,000 K, and this is somewhat lower than the temperatures derived for the rapidly evolving transients from PS1 \citep{PS1}. A possible reason is that these samples are located at higher redshifts and we observe shorter wavelengths in the rest wavelengths ($<$ 3,000 \AA), which can suffer from higher interstellar extinction. Since there is no way to estimate the extinctions in the host galaxies, the derived temperatures are likely to be a lower limit for the intrinsic temperatures.

In Figure $\ref{fig:peakvsz}$, we show absolute magnitudes in the {\it rest} frame (star symbols) by applying $K$-corrections based on the blackbody fit. Since the blackbody fitting is performed in the rest frame, we simply use the flux density at the rest wavelength of each filter to derive the $K$-corrected, absolute magnitude in the rest frame. 
Note that the large differences between the observed $g$-band magnitudes and the $K$-corrected, rest-frame $g$-band magnitudes of HSC17bbaz ($z=1.480$) and HSC17dadp ($z=0.830$) are due to the fact that the observed $g$-band corresponds to the rest wavelength of $< 3000$ \AA\ and the observed flux is largely attenuated there (Figure \ref{fig:BBfit}). The large difference in the $r$-band data of HSC17auls is caused due to the lack of $r$-band data near the $i-$band peak.
\blue{
The estimated photospheric radii are $3 \times 10^{14} - 6 \times 10^{15}$ cm. By assuming that the time to the peak is twice of $t_{1/2, \rm{rise}}$, these radii correspond to the photospheric velocities of 7,000 - 64,000 $\rm{km \ s^{-1}}$.}
As the estimated temperature is likely a lower limit as discussed above, the photospheric radii and velocities tend to be overestimated. Therefore, the true velocities are likely to be $ \sim  10,000  {\rm km \ s^{-1}}$ or lower.

\startlongtable
\begin{deluxetable*}{c|cccc}
    \tablecaption{\green{Photometric properties of the rapidly evolving transients from HSC-SSP Transient Survey}
    \label{tab:RET_lc}}
    \tablehead{
      \colhead{Name} & \colhead{$M_{\rm peak}$\tablenotemark{a}} & \colhead{$t_{1/2,\rm rise}$\tablenotemark{a}} & \colhead{$t_{1/2, \rm decline}$\tablenotemark{a}}  & \colhead{Temperature\tablenotemark{b}}\\
     \colhead{} & \colhead{(mag)} & \colhead{(day)} & \colhead{(day)} & \colhead{(K)}
    } 
    \startdata
    HSC17auls & -17.53$\pm\footnotesize{0.02}$ & 1.37$\pm\footnotesize{0.08}$ & 5.10$\pm\footnotesize{0.90}$ & 17,400$\pm\footnotesize{1,900}$\\
    HSC17bbaz & -19.76$\pm\footnotesize{0.11}$ & 5.11$\pm\footnotesize{0.61}$ & 6.37$\pm\footnotesize{0.06}$ & 8,100$\pm\footnotesize{1,400}$\\
    HSC17bhyl & -18.49$\pm\footnotesize{0.04}$ & 2.38$\pm\footnotesize{0.31}$ & 7.15$\pm\footnotesize{1.95}$ & 15,400$\pm\footnotesize{1,300}$\\
    HSC17btum & -17.35$\pm\footnotesize{0.05}$ & 2.58$\pm\footnotesize{0.54}$ & 15.96$\pm\footnotesize{0.26}$ & 16,900$\pm\footnotesize{3,700}$\\
    HSC17dadp & -18.74$\pm\footnotesize{0.03}$ & 1.88$\pm\footnotesize{0.43}$ & 7.36$\pm\footnotesize{0.29}$ & 8,100$\pm\footnotesize{1,400}$
    \enddata
    \tablenotetext{a}{All the magnitudes and timescales are given for the $i$-band. The definitions of timescales ($t_{1/2, \rm rise}$ and $t_{1/2, \rm decline}$) are described in Section \ref{sec:photometric}.}
    \tablenotetext{b}{\green{Blackbody temperature at the $i$-band peak.}}
\end{deluxetable*}

\section{discussion}
\label{sec:discussion}

\subsection{Host galaxies}
\label{sec:galaxies}

We present the properties of the host galaxies of our rapidly evolving transients in Table \ref{tab:host_galaxies}. 
Star formation rates (SFRs) of the host galaxies of rapidly evolving transients can give a clue to the progenitor scenario for this class of transients. Since multi-band data for the COSMOS field are available from the HSC-SSP survey \citep{hsc_basic}, we can estimate the SFRs based on the multi-color photometry. Here we use the SFRs estimated from the galaxy template fitting code MIZUKI \citep{tanaka_photoz,  HSC_photoz}.
MIZUKI performs a fitting with the theoretical templates parametarized by stellar population, initial mass function, dust attenuation and so on. By the fitting, we can estimate the physical parameters using the galaxy templates like stellar mass and SFR.

\begin{figure}[t]
 \begin{center}
   \includegraphics[width=85mm]{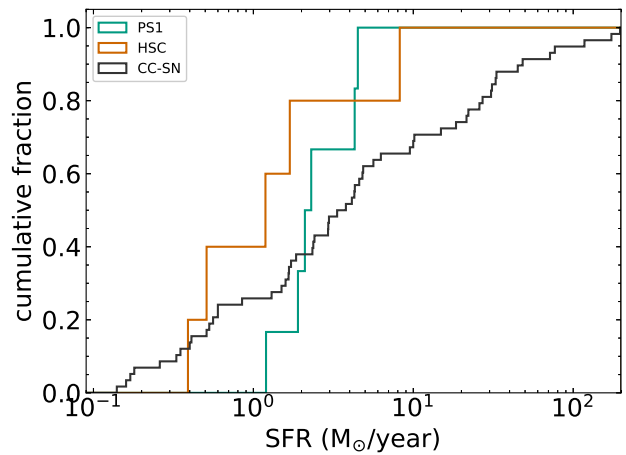}
 \end{center}
 \caption{Cumulative distributions of SFRs of the host galaxies of our samples (red), samples from PS1 (green, \citealt{PS1}) and core-collapse SNe (gray, \citealt{CCSFR}). }
 \label{fig:SFR}
\end{figure}

We find that all the host galaxies of our samples have SFR $\ge$ 0.1 M$_\odot$ yr$^{-1}$. This means that the host galaxies of our rapidly evolving transients are star forming galaxies. In Figure $\ref{fig:SFR}$, we show cumulative distribution of SFRs for the rapidly evolving transients from the HSC-SSP Transient Survey (this paper) and PS1 \citep{PS1} as well as those for normal core-collapse SNe \citep{CCSFR}.
We performed Kolmogorov-Smirnov (K-S) test and that there is no significant evidence ($p =$ 0.22) that the distribution of SFRs are different between core-collapse SNe and the rapidly evolving transients from the HSC-SSP Transient Survey, although the number of the rapidly evolving transients is limited.
This suggests that the progenitors of our rapidly evolving transients may be massive stars.

\blue{We also estimated the offsets of the transients from the centers of their host galaxies. They range from 0.4 to 25.7 kpc. These offsets except HSC17bbaz are within the range of offsets found for core-collapse SNe \citep{prieto08_host_dist}.
}

\begin{table}[t]
    \begin{center}
      \caption{Properties of the host galaxies}
      \begin{tabular}{c || ccc} \hline
        Name & Redshift & SFR & Offset  \\ 
             &          & (M$_\odot$ $\rm{ yr^{-1}}$) & (kpc)\\ \hline
        HSC17auls & 0.339 & 1.19 & 3.73 \\
        HSC17bbaz & 1.480 & 8.23 & 25.69  \\
        HSC17bhyl & 0.750 & 1.69 & 0.43  \\
        HSC17btum & 0.467 & 0.51 & 13.01  \\
        HSC17dadp & 0.830 & 0.39 & 5.27  \\ \hline
     \end{tabular}
  \end{center}
  \label{tab:host_galaxies}
\end{table}

\subsection{Event Rate}
In this section, we estimate the event rate of our rapidly evolving transients. 
A precise measurement of the event rate requires knowledge about luminosity function and light curve shapes of the transients. However, these are not yet clear for rapidly evolving transients. Thus, we derive an approximated event rate by using the following method which is free from assumptions about the light curve properties.
Using index of transient $i$, redshift $z_i$, observing efficiency $\epsilon$, and survey duration $T_i$ (typically $T_i$ $\sim$ 6 months),
the event rate can be estimated as follows:
\begin{eqnarray}
 r =\sum_{i}^N \frac{(1+z_i)}{\epsilon T_i V_{{\rm max},i}}.
  \label{eq:eventrate}
\end{eqnarray}
Here $V_{\rm max,i}$ is the comoving volume for the maximum redshift to which the transient is detectable with our data. We assume $\epsilon =$ 1. Note that since the cadence of the HSC-SSP Transient Survey is about one week and the observations were performed only around new moon phases, the true efficiency should be $\epsilon < 1$. Therefore, the estimated event rate gives a lower bound.

\begin{figure}[t]
 \begin{center}
   \includegraphics[width=85mm]{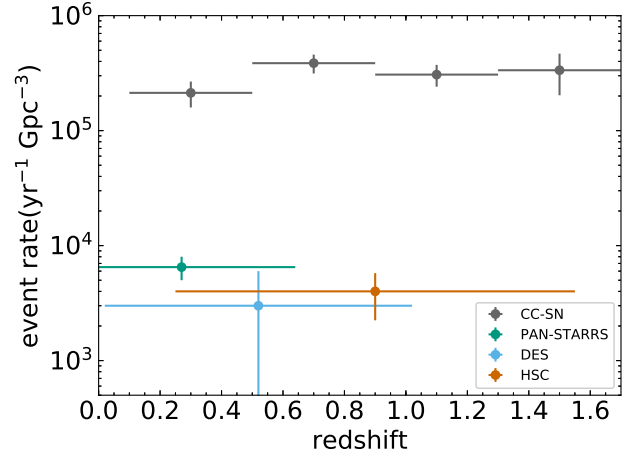}
 \end{center}
 \caption{Event rates of the rapidly evolving transient from the HSC-SSP Transient Survey (red), PS1 (green; \citealt{PS1}), and Dark Energy Survey (blue; \citealt{DES}) compared with the event rate of normal core-collapse SNe (gray; \citealt{II_eventrate}).}
 \label{fig:eventrate}
\end{figure}

\green{The estimated individual event rates for our samples are 1,050, 250, 600, 1,600 and 500 event yr$^{-1}$ Gpc$^{-3}$ for HSC17auls, HSC17bbaz, HSC17bhyl, HSC17btum and HSC17dadp, respectively.} 
By summing up these (Equation $\ref{eq:eventrate}$), the total event rate is estimated to be $\sim$4,000 events year$^{-1}$ Gpc$^{-3}$.
Figure $\ref{fig:eventrate}$ summarizes the event rates of rapidly evolving transients and normal core-collapse SNe. 
Although the number of our samples is limited, the estimated event rate is roughly consistent with that from PS1 (4,800 $\sim$ 8,000 events year$^{-1}$ Gpc$^{-3}$; \citealt{PS1}) and from Dark Energy Survey ($\ge$ 10$^3$ events year$^{-1}$ Gpc$^{-3}$; \citealt{DES}). This event rate corresponds to $\sim$1 $\%$ of core-collapse SNe events at z = 0.7 (3.86 $\times$ 10$^5$ events year$^{-1}$ Gpc$^{-3}$, \citealt{II_eventrate}).
Also, the event rate of our rapidly evolving transients is about 50 times larger than that of SLSN at z $\sim$ 1 (91 events year$^{-1}$ Gpc$^{-3}$; \citealt{SLSN_eventrate}), which indicates that rapidly evolving transients arise from more common stellar evolutionary paths than SLSNe.

\subsection{Power Source}

\begin{figure}[tbp]
 \begin{center}
   \includegraphics[width=85mm]{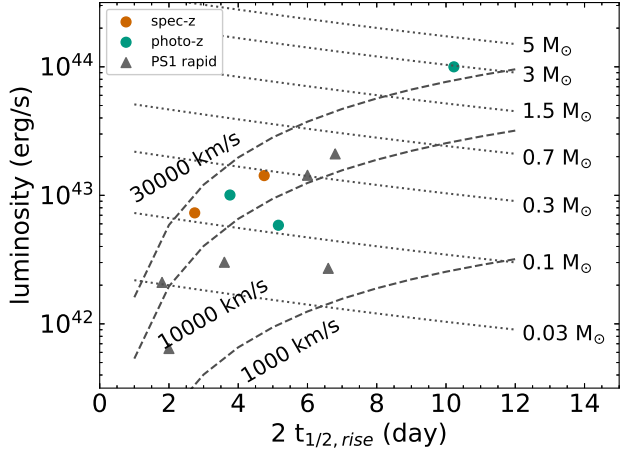}
 \end{center}
 \caption{2 $t_{1/2,\rm rise}$ vs observed peak luminosity of our rapidly evolving transients and the rapidly evolving transients from PS1. Lines represent the correlation of $t_{1/2,\rm rise}$ and observed peak luminosity at given parameters assuming mass of $^{56}$Ni is equal to the mass of ejecta.}
 \label{fig:powersource2}
\end{figure}

 In Figure $\ref{fig:powersource2}$, we show the relation between the rise times and the  peak luminosities of the rapidly evolving transients from the HSC-SSP Transient Survey and PS1 \citep{PS1}.
 For the rise time ($t_{\rm rise}$), we adopt $t_{\rm rise} = 2 \ t_{\rm 1/2, rise}$.
 This gives a sound approximation for the rise time: the mean $t_{\rm 1/2, rise}$ of our SN Ia samples is $\sim$10 days while a typical rise time of SNe Ia is known to be about 20 days \citep{Ia_rise_time}, although the luminosity source of rapidly evolving transients may be different from that of SNe Ia.
 Figure \ref{fig:powersource2} shows that our rapidly evolving transients share a similar parameter space with the PS1 objects.
 As shown by \citet{PS1}, these rapidly evolving transients are as luminous as normal SNe Ia, but have shorter timescales than SNe Ia.

Motivated by similar analysis by \citet{PS1} and \citet{arcavi_rapid}, we discuss the power source
of rapidly evolving transients using the phase diagram (Figure \ref{fig:powersource2}).
The location in the phase diagram has several implications as discussed below.
The dotted lines show the radioactive decay luminosity for given masses of $^{56}$Ni \citep{Ni_decay}:

\begin{eqnarray}
  L = \left[L_{\rm Ni} \exp \left( {-\frac{2 \ t_{1/2,\rm rise}}{\tau_{\rm Ni}}} \right) + L_{\rm Co} \exp \left( {- \frac{2 \ t_{1/2,\rm rise}}{\tau_{\rm Co}}} \right) \right] \nonumber \\ \left( \frac{M_{\rm Ni}}{0.1 \rm M_\odot} \right),
  \label{eq:Ni-mass}
\end{eqnarray}
where $L_{\rm Ni}=6.5 \times 10^{42} \ {\rm erg\ s^{-1}}$, $L_{\rm Co}=1.5 \times 10^{42}\ {\rm erg\ s^{-1}}$, $\tau_{\rm Ni}=8.8$ days, and $\tau_{\rm Co}=111$ days.
The rising timescale $t_{1/2, \rm rise}$ and the luminosity $L$ of the samples are calculated using the $i$-band light curves and blackbody fitting to the rest-frame SED, respectively (Section \ref{sec:photometric}). 

The rise time can be linked to the ejecta mass, velocity and opacity $\kappa$ \citep{1982arnett}:
\begin{eqnarray}
  t_{\rm rise} \sim 15 \ {\rm days} \left( \frac{\kappa}{0.1 \ {\rm cm}^{2}\ {\rm g}^{-1}} \right)^{1/2}  \left( \frac{M_{\rm ejecta}}{1.4 \ {\rm M}_{\odot}} \right) ^{1/2} \nonumber \\ \left( \frac{v}{10^4 \ \rm km\ s^{-1}} \right) ^{-1/2}.
   \label{eq:ejecta-mass}
\end{eqnarray}
Here we simply assume one-zone ejecta and therefore, the exact rise time is somewhat uncertain, depending on the density profile of the ejecta and distribution of $^{56}$Ni in the ejecta. Then, for a given velocity and a rise time, the ejecta mass can be roughly estimated. The dashed lines in Figure \ref{fig:powersource2} show the maximum luminosities from the decay of $^{56}$Ni, by assuming $M_{\rm ejecta} = M_{\rm Ni}$.
For a given velocity, the luminosity of the objects located at upper side of the diagram cannot be explained by the decay of $^{56}$Ni since it requires $ M_{\rm Ni} > M_{\rm ejecta}$.
By assuming typical velocity of $v \sim 10,000 \ {\rm km\ s^{-1}}$,
more than half of our rapidly evolving transients are located above this maximum luminosity.
This indicates that some of the rapidly evolving transients may be difficult to be explained only by the radioactive decay of $^{56}$Ni.

Since the rapidly evolving transients have large varieties in their luminosities and timescales, their power sources may not be explained by a single scenario. Some events seem to be explained well by shock breakout a from progenitor star \citep{PS1}, or from CSM (\citealt{PS1}, \citealt{KSN2015K}). 
Although deep systematic transient surveys such as PS1, DES, and Subaru/HSC can ensure the discovery of a certain number of rapidly evolving transients, spectroscopy is still challenging due to the faintness as well as difficulties in real-time classification. 
This situation will become more severe in the era of Large Synoptic Survey Telescope \citep{lsst}.
Classification with limited photometric data and rapid spectroscopic response are therefore essential to obtain more information on the progenitor scenarios and to understand the nature of rapidly evolving transients.

\section{conclusions}
\label{sec:conclusions}
We performed a systematic search for rapidly evolving transients using the data taken with the HSC-SSP Transient Survey. We identified 5 rapidly evolving transients. These transients have a wide range of the redshift (0.3 $\le$ $z$ $\le$ 1.5) and peak brightness ($-$17 $\ge$ M${\rm _i}$ $\ge$ $-$20). The timescales of our samples are characterized by a rapid rise (t$_{1/2,\rm rise}$ $\le$ 6 days) and slightly slower decline. Overall properties are similar to the rapidly evolving transients identified with PS1. However, the estimated temperatures for our rapidly evolving transients extend to the lower temperature ($\sim$ 8,000 K). This may reflect the fact that the HSC-SSP Transient Survey probes shorter rest wavelengths, which suffer from larger interstellar extinction.

We estimated the total luminosities and timescales  of our samples.
We found that some of the objects are difficult to be explained only by the radioactive decay of $^{56}$Ni due to their high luminosities and short timescales. This fact implies the necessity of other power sources such as shock breakout from CSM.

All the host galaxies of the rapidly evolving transients are star forming galaxies, and thus their progenitors are likely to be massive stars. The event rate is estimated to be $\sim$4,000 events year$^{-1}$ Gpc$^{-3}$, which is $\sim$1 $\%$ of that of core-collapse SNe. The event rate is $\sim$50 times larger than that of SLSNe, which indicates that rapidly evolving transients arise from more common evolutionary paths than SLSNe.

\begin{longtable}{ccccc}
  \caption{\greenn{Log of photometric observations for our samples}}\label{tab:obs_log} 
    \\
    \hline              
    Name & MJD & Mag$^a$ & Error$^b$ & Band\\ 
    \endfirsthead
    \hline
    Name & MJD & Mag$^a$ & Error$^b$ & Band\\ \hline 
    \endhead
    \hline
    \multicolumn{5}{l}{$^a$ 3$\sigma$ upper limit is given for non detection.}\\
    \multicolumn{5}{l}{$^b$ 1$\sigma$ error is presented.} \\
    \endfoot
  \hline
  \hline
      HSC17auls & 57715.54 & $<$25.72 & -- & $z$ \\ 
     & 57717.57 & $<$25.61 & -- & $g$ \\ 
     & 57717.62 & $<$26.11 & -- & $i$ \\ 
     & 57720.60 & $<$26.75 & -- & $r$ \\ 
     & 57721.55 & $<$25.81 & -- & $i$ \\ 
     & 57721.60 & $<$25.40 & -- & $z$ \\ 
     & 57745.56 & $<$25.55 & -- & $z$ \\ 
     & 57747.53 & $<$26.10 & -- & $r$ \\ 
     & 57747.62 & $<$25.52 & -- & $i$ \\ 
     & 57755.45 & 23.61 &  0.06 & $z$ \\
     & 57755.51 & 23.48 &  0.02 & $i$ \\
     & 57755.61 & 23.01 &  0.01 & $g$ \\
     & 57774.50 & 25.14 &  0.08 & $z$ \\
     & 57776.40 & $<$26.56 & -- & $r$ \\ 
     & 57776.55 & 25.80 &  0.13 & $i$ \\
     & 57778.45 & $<$25.64 & -- & $g$ \\ 
     & 57783.44 & $<$26.31 & -- & $i$ \\ 
     & 57783.55 & $<$26.26 & -- & $z$ \\ 
     & 57785.40 & $<$26.12 & -- & $g$ \\ 
     & 57786.46 & $<$26.71 & -- & $r$ \\ 
     & 57786.60 & $<$25.91 & -- & $i$ \\ 
     & 57805.37 & $<$26.35 & -- & $z$ \\ 
     & 57807.37 & $<$25.80 & -- & $g$ \\ 
     & 57807.48 & $<$26.30 & -- & $r$ \\ 
     & 57809.41 & $<$26.11 & -- & $i$ \\ 
     & 57816.30 & $<$25.89 & -- & $z$ \\ 
     & 57816.47 & $<$26.68 & -- & $i$ \\ 
     & 57818.51 & $<$26.53 & -- & $r$ \\ 
     & 57834.31 & $<$26.74 & -- & $g$ \\ 
     & 57834.43 & $<$26.37 & -- & $z$ \\ 
     & 57835.26 & $<$26.35 & -- & $i$ \\ 
     & 57837.27 & $<$25.95 & -- & $r$ \\ 
     & 57841.29 & $<$26.49 & -- & $g$ \\ 
     & 57841.41 & $<$25.89 & -- & $z$ \\ 
     & 57842.27 & $<$26.17 & -- & $i$ \\ 
     & 57844.33 & $<$25.74 & -- & $r$ \\ 
     & 57866.25 & $<$26.19 & -- & $r$ \\ 
     & 57866.36 & $<$25.57 & -- & $z$ \\ 
     & 57869.27 & $<$25.33 & -- & $i$ \\ 
     & 57869.33 & $<$25.53 & -- & $g$ \\ 
     & 57870.35 & $<$26.21 & -- & $i$ \\ 
     & 57872.26 & $<$25.69 & -- & $z$ \\ 
     & 57924.28 & $<$23.92 & -- & $z$ \\ 
     & & & & \\
    HSC17bbaz & 57745.53 & $<$25.46 & -- & $z$ \\ 
     & 57747.53 & $<$25.93 & -- & $r$ \\ 
     & 57747.61 & $<$25.50 & -- & $i$ \\ 
     & 57755.53 & 25.18 &  0.15 & $i$ \\
     & 57755.61 & $<$26.71 & -- & $g$ \\ 
     & 57774.53 & 23.86 &  0.04 & $z$ \\
     & 57776.40 & 24.93 &  0.09 & $r$ \\
     & 57776.52 & 23.93 &  0.05 & $i$ \\
     & 57778.46 & $<$25.70 & -- & $g$ \\ 
     & 57783.45 & 24.19 &  0.09 & $i$ \\
     & 57783.56 & 23.82 &  0.05 & $z$ \\
     & 57785.40 & $<$26.32 & -- & $g$ \\ 
     & 57786.46 & 25.25 &  0.12 & $r$ \\
     & 57786.57 & 24.17 &  0.05 & $i$ \\
     & 57805.37 & 24.93 &  0.18 & $z$ \\
     & 57807.37 & $<$26.19 & -- & $g$ \\ 
     & 57807.49 & $<$26.29 & -- & $r$ \\ 
     & 57809.41 & $<$25.03 & -- & $i$ \\ 
     & 57816.30 & $<$25.58 & -- & $z$ \\ 
     & 57816.48 & $<$26.31 & -- & $i$ \\ 
     & 57818.52 & $<$26.15 & -- & $r$ \\ 
     & 57834.31 & $<$26.71 & -- & $g$ \\ 
     & 57834.45 & $<$25.50 & -- & $z$ \\ 
     & 57835.26 & $<$25.93 & -- & $i$ \\ 
     & 57837.27 & $<$26.16 & -- & $r$ \\ 
     & 57841.29 & $<$26.28 & -- & $g$ \\ 
     & & & & \\
    HSC17bhyl & 57715.54 & $<$25.98 & -- & $z$ \\ 
     & 57717.57 & $<$26.19 & -- & $g$ \\ 
     & 57717.62 & $<$26.58 & -- & $i$ \\ 
     & 57720.60 & $<$27.08 & -- & $r$ \\ 
     & 57721.55 & $<$26.46 & -- & $i$ \\ 
     & 57721.61 & $<$25.79 & -- & $z$ \\ 
     & 57745.57 & $<$25.72 & -- & $z$ \\ 
     & 57747.52 & $<$26.30 & -- & $r$ \\ 
     & 57747.63 & $<$26.02 & -- & $i$ \\ 
     & 57755.44 & $<$25.69 & -- & $z$ \\ 
     & 57755.51 & $<$26.91 & -- & $i$ \\ 
     & 57755.61 & $<$27.24 & -- & $g$ \\ 
     & 57774.50 & 24.78 &  0.05 & $z$ \\
     & 57776.41 & 24.10 &  0.03 & $r$ \\
     & 57776.55 & 24.29 &  0.03 & $i$ \\
     & 57778.45 & 24.09 &  0.07 & $g$ \\
     & 57779.52 & 24.41 &  0.19 & $z$ \\
     & 57783.43 & 24.46 &  0.06 & $i$ \\
     & 57783.55 & 24.62 &  0.08 & $z$ \\
     & 57785.40 & 25.63 &  0.11 & $g$ \\
     & 57786.45 & 25.16 &  0.05 & $r$ \\
     & 57786.58 & 24.73 &  0.08 & $i$ \\
     & 57805.37 & $<$26.20 & -- & $z$ \\ 
     & 57807.37 & $<$26.62 & -- & $g$ \\ 
     & 57807.48 & $<$26.76 & -- & $r$ \\ 
     & 57809.41 & $<$26.42 & -- & $i$ \\ 
     & 57816.31 & $<$26.36 & -- & $z$ \\ 
     & 57816.47 & 26.23 &  0.17 & $i$ \\
     & 57818.52 & $<$26.72 & -- & $r$ \\ 
     & 57834.33 & $<$26.90 & -- & $g$ \\ 
     & 57834.44 & $<$26.42 & -- & $z$ \\ 
     & 57835.27 & $<$26.32 & -- & $i$ \\ 
     & 57837.27 & $<$26.34 & -- & $r$ \\ 
     & 57841.29 & $<$27.04 & -- & $g$ \\ 
     & 57841.40 & $<$26.10 & -- & $z$ \\ 
     & 57842.27 & $<$26.15 & -- & $i$ \\ 
     & 57844.33 & $<$26.73 & -- & $r$ \\ 
     & 57866.25 & $<$26.82 & -- & $r$ \\ 
     & 57866.36 & $<$25.99 & -- & $z$ \\ 
     & 57869.27 & $<$26.28 & -- & $i$ \\ 
     & 57870.35 & $<$26.49 & -- & $i$ \\ 
     & 57872.26 & $<$25.92 & -- & $z$ \\ 
     & 57924.28 & $<$24.41 & -- & $z$ \\ 
     & & & & \\
    HSC17btum & 57715.55 & $<$25.23 & -- & $z$ \\ 
     & 57717.57 & $<$25.40 & -- & $g$ \\ 
     & 57717.62 & $<$26.38 & -- & $i$ \\ 
     & 57720.60 & $<$26.74 & -- & $r$ \\ 
     & 57721.54 & $<$25.25 & -- & $i$ \\ 
     & 57721.61 & $<$25.00 & -- & $z$ \\ 
     & 57745.54 & $<$25.39 & -- & $z$ \\ 
     & 57747.51 & $<$26.24 & -- & $r$ \\ 
     & 57747.61 & $<$25.82 & -- & $i$ \\ 
     & 57755.45 & $<$25.46 & -- & $z$ \\ 
     & 57755.52 & $<$26.00 & -- & $i$ \\ 
     & 57755.62 & $<$26.92 & -- & $g$ \\ 
     & 57774.50 & $<$26.63 & -- & $z$ \\ 
     & 57776.40 & $<$26.88 & -- & $r$ \\ 
     & 57776.53 & $<$26.89 & -- & $i$ \\ 
     & 57778.45 & $<$25.94 & -- & $g$ \\ 
     & 57783.44 & 24.84 &  0.11 & $i$ \\
     & 57783.56 & 25.22 &  0.11 & $z$ \\
     & 57785.39 & 24.00 &  0.02 & $g$ \\
     & 57786.46 & 24.00 &  0.03 & $r$ \\
     & 57786.59 & 24.38 &  0.05 & $i$ \\
     & 57805.36 & 24.52 &  0.14 & $z$ \\
     & 57807.37 & 25.08 &  0.10 & $g$ \\
     & 57807.48 & 24.65 &  0.09 & $r$ \\
     & 57809.41 & 25.06 &  0.14 & $i$ \\
     & 57816.31 & $<$25.65 & -- & $z$ \\ 
     & 57816.47 & 26.02 &  0.19 & $i$ \\
     & 57818.51 & $<$26.73 & -- & $r$ \\ 
     & 57834.32 & $<$27.06 & -- & $g$ \\ 
     & 57834.43 & $<$26.08 & -- & $z$ \\ 
     & 57835.27 & $<$26.61 & -- & $i$ \\ 
     & 57837.27 & $<$26.59 & -- & $r$ \\ 
     & 57841.29 & $<$26.42 & -- & $g$ \\ 
     & 57841.41 & $<$25.83 & -- & $z$ \\ 
     & 57842.27 & $<$25.92 & -- & $i$ \\ 
     & 57844.33 & $<$26.10 & -- & $r$ \\ 
     & 57866.36 & $<$25.42 & -- & $z$ \\ 
     & 57872.27 & $<$25.28 & -- & $z$ \\ 
     & 57924.28 & $<$23.72 & -- & $z$ \\ 
     & & & & \\
    HSC17dadp & 57715.55 & $<$26.07 & -- & $z$ \\ 
     & 57717.57 & $<$26.13 & -- & $g$ \\ 
     & 57717.62 & $<$26.60 & -- & $i$ \\ 
     & 57720.60 & $<$26.96 & -- & $r$ \\ 
     & 57721.55 & $<$26.32 & -- & $i$ \\ 
     & 57721.60 & $<$26.15 & -- & $z$ \\ 
     & 57745.56 & $<$25.93 & -- & $z$ \\ 
     & 57747.53 & $<$26.66 & -- & $r$ \\ 
     & 57747.62 & $<$26.22 & -- & $i$ \\ 
     & 57755.45 & $<$26.04 & -- & $z$ \\ 
     & 57755.51 & $<$27.01 & -- & $i$ \\ 
     & 57755.61 & $<$27.66 & -- & $g$ \\ 
     & 57774.51 & $<$27.23 & -- & $z$ \\ 
     & 57776.40 & $<$27.11 & -- & $r$ \\ 
     & 57776.54 & $<$26.96 & -- & $i$ \\ 
     & 57778.45 & $<$26.47 & -- & $g$ \\ 
     & 57779.52 & $<$25.17 & -- & $z$ \\ 
     & 57783.43 & $<$26.42 & -- & $i$ \\ 
     & 57783.55 & $<$26.26 & -- & $z$ \\ 
     & 57785.40 & $<$27.17 & -- & $g$ \\ 
     & 57786.45 & $<$26.98 & -- & $r$ \\ 
     & 57786.58 & $<$26.72 & -- & $i$ \\ 
     & 57805.37 & $<$26.37 & -- & $z$ \\ 
     & 57807.37 & $<$27.00 & -- & $g$ \\ 
     & 57807.49 & $<$27.08 & -- & $r$ \\ 
     & 57809.41 & 25.81 &  0.19 & $i$ \\
     & 57816.31 & 24.36 &  0.05 & $z$ \\
     & 57816.47 & 24.26 &  0.03 & $i$ \\
     & 57818.51 & 24.30 &  0.03 & $r$ \\
     & 57834.31 & $<$27.42 & -- & $g$ \\ 
     & 57834.44 & 24.87 &  0.08 & $z$ \\
     & 57835.26 & 25.31 &  0.13 & $i$ \\
     & 57837.27 & $<$26.57 & -- & $r$ \\ 
     & 57841.29 & $<$26.94 & -- & $g$ \\ 
     & 57841.41 & 25.27 &  0.17 & $z$ \\
     & 57842.27 & $<$26.62 & -- & $i$ \\ 
     & 57844.33 & $<$26.68 & -- & $r$ \\ 
     & 57866.25 & $<$26.63 & -- & $r$ \\ 
     & 57866.36 & $<$25.69 & -- & $z$ \\ 
     & 57869.27 & $<$26.10 & -- & $i$ \\ 
     & 57869.33 & $<$26.85 & -- & $g$ \\ 
     & 57870.35 & $<$26.82 & -- & $i$ \\ 
     & 57872.26 & $<$25.69 & -- & $z$ \\ 
     & 57924.28 & $<$24.33 & -- & $z$ 
\end{longtable}

\acknowledgments
The Hyper Suprime-Cam (HSC) collaboration includes the astronomical communities of Japan, Taiwan, and Princeton University. The HSC instrumentation and software were developed by the National Astronomical Observatory of Japan (NAOJ), the Kavli Institute for the Physics and Mathematics of the Universe (Kavli IPMU), the University of Tokyo, the High Energy Accelerator Research Organization (KEK), the Academia Sinica Institute for Astronomy and Astrophysics in Taiwan (ASIAA), and Princeton University. Funding was contributed by the FIRST program from the Japanese Cabinet Office, the Ministry of Education, Culture, Sports, Science and Technology (MEXT), the Japan Society for the Promotion of Science (JSPS), the Japan Science and Technology Agency (JST), the Toray Science Foundation, NAOJ, Kavli IPMU, KEK, ASIAA, and Princeton University. The Pan-STARRS1 Surveys have been made possible through contributions by the Institute for Astronomy, the University of Hawaii, the Pan-STARRS Project Office, the Max-Planck Society and its participating institutes, the Max Planck Institute for Astronomy, Heidelberg and the Max Planck Institute for Extraterrestrial Physics, Garching, Johns Hopkins University, Durham University, the University of Edinburgh, Queen’s University Belfast, the Harvard-Smithsonian Center for Astrophysics, the Las Cumbres Observatory Global Telescope Network Incorporated, the National Central University of Taiwan, the Space Telescope Science Institute, the National Aeronautics and Space Administration (under Grant No. NNX08AR22G issued through the Planetary Science Division of the NASA Science Mission Directorate), the National Science Foundation (under Grant No. AST-1238877), the University of Maryland, and Eotvos Lorand University (ELTE). This study used software developed for the Large Synoptic Survey Telescope (LSST). We thank the LSST Project for making their code available as free software at http://dm.lsst.org. 
M. T. is supported by the Grant-in-Aid for Scientific Research programs of JSPS (15H02075, 16H02183, 19H00694) and MEXT (17H06363).
K. M. is supported by the Grant-in-Aid for Scientific Research programs of JSPS (18H0455, 18H05223, 17H02864).

\blue{
\software{sncosmo \citep{barbary16}, SNANA  \citep{kessler09}, MIZUKI \citep{tanaka_photoz,  HSC_photoz}, MLZ \citep{MLZ}, EPHOR \footnote[1]{\url{https://hsc-release.mtk.nao.ac.jp/doc/index.php/photometric-redshifts/}} and Franken-z (Speagle et al. in prep)}.
}

\clearpage
\bibliography{HSC_COSMOS_rapid}
\bibliographystyle{aasjournal}

\end{document}